\documentclass[letterpaper,twocolumn,prl,aps,superscriptaddress,amsmath,amssymb,floatfix]{revtex4-1}
\usepackage{mathptmx}
\usepackage[latin9]{inputenc}
\usepackage{color}
\usepackage{verbatim}
\usepackage{float}
\usepackage{amsmath}
\usepackage{amssymb}
\usepackage{graphicx}
\usepackage{esint}

\usepackage[unicode=true,
 bookmarks=true,bookmarksnumbered=false,bookmarksopen=false,
 breaklinks=false,pdfborder={0 0 1},backref=false,colorlinks=true]
 {hyperref}
\hypersetup{
 linkcolor=magenta,urlcolor=blue,citecolor=blue,pdfstartview={FitH},hyperfootnotes=false}

\makeatletter

\usepackage{textcomp}
\usepackage{epstopdf}

\pdfpageheight\paperheight
\pdfpagewidth\paperwidth

\@ifundefined{textcolor}{}{%
 \definecolor{BLACK}{gray}{0}
 \definecolor{WHITE}{gray}{1}
 \definecolor{RED}{rgb}{1,0,0}
 \definecolor{GREEN}{rgb}{0,1,0}
 \definecolor{BLUE}{rgb}{0,0,1}
 \definecolor{CYAN}{cmyk}{1,0,0,0}
 \definecolor{MAGENTA}{cmyk}{0,1,0,0}
 \definecolor{YELLOW}{cmyk}{0,0,1,0}
}

\usepackage{xcolor}
\usepackage{soul}
\definecolor{blue}{rgb}{0,0,1}
\definecolor{red}{rgb}{1,0,0}
\definecolor{green}{rgb}{0,1,0}

\usepackage{soul}

\makeatother

\begin{document}
\title{Beating the break-even point with a discrete-variable-encoded logical qubit}

\author{Zhongchu Ni}
\affiliation{Shenzhen Institute for Quantum Science and Engineering, Southern University of Science and Technology, Shenzhen 518055, China}
\affiliation{International Quantum Academy, Shenzhen 518048, China}
\affiliation{Guangdong Provincial Key Laboratory of Quantum Science and Engineering, Southern University of Science and Technology, Shenzhen 518055, China}
\affiliation{Department of Physics, Southern University of Science and Technology, Shenzhen 518055, China}

\author{Sai Li}
\author{Xiaowei Deng}
\author{Yanyan Cai}
\author{Libo Zhang}
\affiliation{Shenzhen Institute for Quantum Science and Engineering, Southern University of Science and Technology, Shenzhen 518055, China}
\affiliation{International Quantum Academy, Shenzhen 518048, China}
\affiliation{Guangdong Provincial Key Laboratory of Quantum Science and Engineering, Southern University of Science and Technology, Shenzhen 518055, China}

\author{Weiting Wang}
\affiliation{Center for Quantum Information, Institute for Interdisciplinary Information Sciences, Tsinghua University, Beijing 100084, China}

\author{Zhen-Biao Yang}
\affiliation{Fujian Key Laboratory of Quantum Information and Quantum Optics, College of Physics and Information Engineering, Fuzhou University, Fuzhou, Fujian 350108, China}

\author{Haifeng Yu}
\affiliation{Beijing Academy of Quantum Information Sciences, Beijing 100193, China}

\author{Fei Yan}
\author{Song Liu}
\affiliation{Shenzhen Institute for Quantum Science and Engineering, Southern University of Science and Technology, Shenzhen 518055, China}
\affiliation{International Quantum Academy, Shenzhen 518048, China}
\affiliation{Guangdong Provincial Key Laboratory of Quantum Science and Engineering, Southern University of Science and Technology, Shenzhen 518055, China}

\author{Chang-Ling Zou}
\affiliation{CAS Key Laboratory of Quantum Information, University of Science and Technology of China, Hefei, Anhui 230026, China}
\author{Luyan Sun}
\email{luyansun@tsinghua.edu.cn}
\affiliation{Center for Quantum Information, Institute for Interdisciplinary Information Sciences, Tsinghua University, Beijing 100084, China}

\author{Shi-Biao Zheng}
\email{t96034@fzu.edu.cn}
\affiliation{Fujian Key Laboratory of Quantum Information and Quantum Optics, College of Physics and Information Engineering, Fuzhou University, Fuzhou, Fujian 350108, China}

\author{Yuan Xu}
\email{xuy5@sustech.edu.cn}
\affiliation{Shenzhen Institute for Quantum Science and Engineering, Southern University of Science and Technology, Shenzhen 518055, China}
\affiliation{International Quantum Academy, Shenzhen 518048, China}
\affiliation{Guangdong Provincial Key Laboratory of Quantum Science and Engineering, Southern University of Science and Technology, Shenzhen 518055, China}

\author{Dapeng Yu}
\email{yudp@sustech.edu.cn}
\affiliation{Shenzhen Institute for Quantum Science and Engineering, Southern University of Science and Technology, Shenzhen 518055, China}
\affiliation{International Quantum Academy, Shenzhen 518048, China}
\affiliation{Guangdong Provincial Key Laboratory of Quantum Science and Engineering, Southern University of Science and Technology, Shenzhen 518055, China}
\affiliation{Department of Physics, Southern University of Science and Technology, Shenzhen 518055, China}

\begin{abstract}
\textbf{Quantum error correction (QEC) aims to protect logical qubits from noises by utilizing the redundancy of a large Hilbert space, where an error, once it occurs, can be detected and corrected in real time. In most QEC codes, a logical qubit is encoded in some discrete variables, e.g., photon numbers. Such encoding schemes make the codewords orthogonal, so that the encoded quantum information can be unambiguously extracted after processing. Based on such discrete-variable encodings, repetitive QEC demonstrations have been reported on various platforms, but there the lifetime of the encoded logical qubit is still shorter than that of the best available physical qubit in the entire system, which represents a break-even point that needs to be surpassed for any QEC code to be of practical use. Here we demonstrate a QEC procedure with a logical qubit encoded in photon-number states of a microwave cavity, dispersively coupled to an ancilla superconducting qubit. By applying a pulse featuring a tailored frequency comb to the ancilla, we can repetitively extract the error syndrome with high fidelity and perform error correction with feedback control accordingly, thereby exceeding the break-even point by about 16\% lifetime enhancement. Our work illustrates the potential of the hardware-efficient discrete-variable QEC codes towards a reliable quantum information processor.
}
\end{abstract}
\maketitle
\vskip 0.5cm

One of the main obstacles for building a quantum computer is environmentally-induced decoherence, which destroys the quantum information stored in the qubits. The errors caused by decoherence can be corrected by repetitive application of a quantum error correction (QEC) procedure, where the logical qubit is encoded in a high-dimensional Hilbert space, such that different errors project the system into different orthogonal subspaces and thus can be unambiguously identified and corrected without disturbing the stored quantum information. In conventional QEC schemes~\cite{nielsen2010, terhal2015}, the codewords of a logical qubit are respectively formed by two highly symmetric entangled states of multiple physical qubits encoded with some discrete variables. The past two decades have witnessed remarkable advances in experimental demonstrations of this kind of QEC code in different systems, including nuclear spins~\cite{cory1998, knill2001}, nitrogen-vacancy centers in diamond~\cite{waldherr2014, abobeih2022}, trapped ions~\cite{chiaverini2004, schindler2011, egan2021, ryan2021, postler2022}, photonic qubits~\cite{yao2012}, silicon spin qubits~\cite{takeda2022}, and superconducting circuits~\cite{reed2012, kelly2015, corcoles2015, Chen2021,krinner2022, zhao2022, acharya2022}. However, in these experiments, the lifetime of the logical qubit still needs to be significantly extended to reach that of the best available physical component, which is regarded as the break-even point for judging whether or not a QEC code can benefit quantum information storage and processing.

An alternative QEC encoding scheme is to employ the large space of an oscillator, which can be used to encode either a continuous-variable or discrete-variable qubit~\cite{cai2021, joshi2021, fluhmann2019, campagne2020, gertler2021}. Both types of codes can tolerate errors due to loss and gain of energy quanta, enabling QEC to be performed in a hardware-efficient way. Breakthrough QEC demonstrations have been reported in a circuit quantum electrodynamics (QED) system~\cite{ofek2016}, where the break-even point was exceeded by distributing the quantum information over an infinite-dimensional Hilbert space, but realized with logical qubits bearing two non-orthogonal codewords. This inherent restriction can be overcome with discrete-variable encoding schemes, where the codewords of a logical qubit are encoded with mutually orthogonal Fock states of an oscillator. This feature, together with their intrinsic compatibility with error-correctable gates~\cite{ma2020, reinhold2020}, as well as with their usefulness for logically connecting modules in a quantum network~\cite{chou2018}, makes such discrete-variable qubits promising in fault-tolerant quantum computation. These advantages can be turned into practical benefits in real quantum information processing only when the lifetime of the encoded logical qubits is extended beyond the break-even point, which, however, remains an elusive task, although enduring efforts have been made towards this goal~\cite{hu2019, gertler2021}.

\begin{figure*}
    \includegraphics{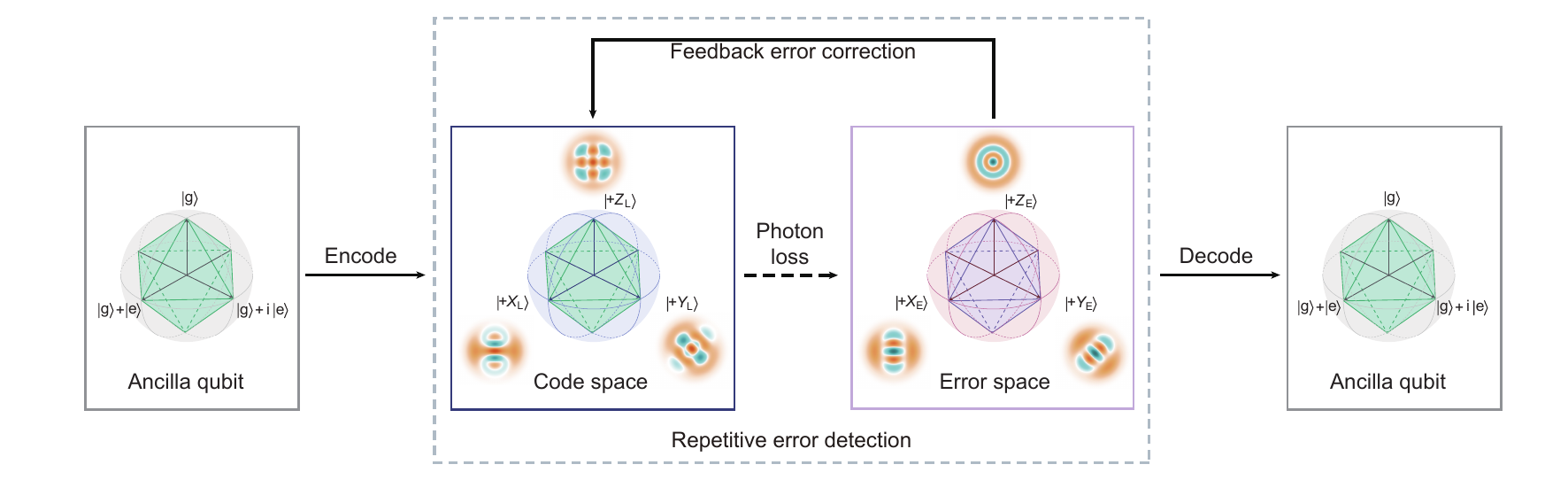}
    \caption{\textbf{Schematic of the QEC procedure with the lowest-order binomially encoded logical qubit.} The ancilla qubit is first encoded to the logical qubit in an oscillator with \{$\left\vert 0_\mathrm{L}\right\rangle = \left( \left\vert 0\right\rangle +\left\vert 4\right\rangle \right) /\sqrt{2}$, $\left\vert 1_\mathrm{L}\right\rangle = \left\vert 2\right\rangle$\}. Once a single-photon-jump error occurs, the logical qubit state falls out of the code space to the error space with the basis states: \{$\left\vert 0_\mathrm{E}\right\rangle = \left\vert 3\right\rangle, \left\vert 1_\mathrm{E}\right\rangle = \left\vert 1\right\rangle $\}. After repetitive error detecting and correcting, the logical qubit state is protected against single-photon-jump errors. Finally, quantum state is decoded back to the ancilla for a final state characterization. The cardinal point states in the Bloch spheres of the code and error spaces are defined as $\vert + Z_\mathrm{L(E)}\rangle = \vert 0_\mathrm{L(E)}\rangle$, $\vert + X_\mathrm{L(E)}\rangle = (\vert 0_\mathrm{L(E)}\rangle + \vert 1_\mathrm{L(E)}\rangle )/\sqrt{2}$, and $\vert + Y_\mathrm{L(E)}\rangle = (\vert 0_\mathrm{L(E)}\rangle + i \vert 1_\mathrm{L(E)}\rangle )/\sqrt{2}$, respectively.}

    \label{fig1}
\end{figure*}

Here, we demonstrate the exceeding of the QEC break-even point by real-time feedback correction for a discrete-variable photonic qubit in a microwave cavity, whose codewords remain mutually orthogonal and can be unambiguously discriminated. The dominate error, single photon loss, of the logical qubit is mapped to the state of a Josephson-junction based nonlinear oscillator that is dispersively coupled to the cavity and serves as an ancilla qubit, realized with a continuous pulse involving an ingeniously tailored comb of frequency components. As the driving frequencies aim at the error space where a photon loss event occurs, perturbations on the logical qubit are highly suppressed when it remains in the encoded logical space. 
Another intrinsic advantage of this error syndrome detection is that the continuous driving protects the system from the ancilla's dephasing noise. We demonstrate this procedure with the lowest-order binomial code and extend the stored quantum information lifetime 16\% longer than the best physical qubit, encoded in the two lowest Fock states and referred to as the Fock qubit. A more important characteristic associated with this error-detecting procedure is that neither the logical nor the error space needs to have a definite parity, which allows the implementation of QEC codes that can tolerate losses of more than one photon.

\begin{figure}
    \includegraphics{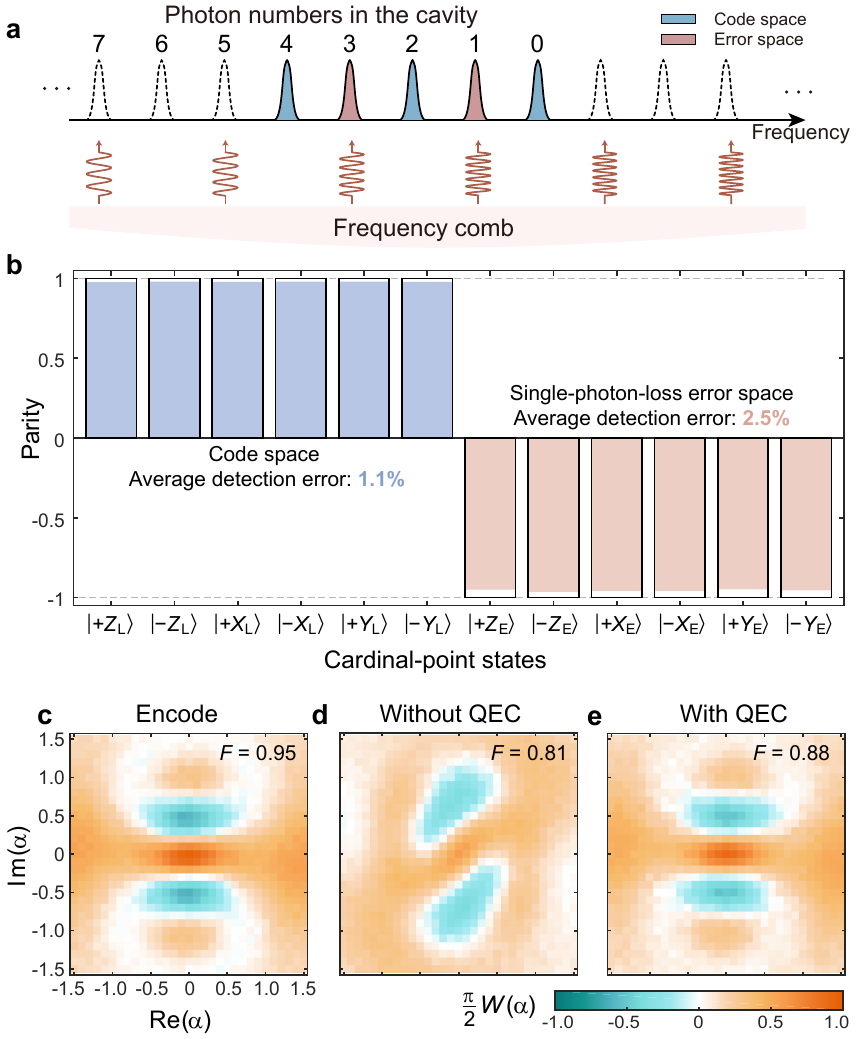}
    \caption{\textbf{Frequency comb control to measure the error syndrome.}
	\textbf{a} The frequency comb control is realized by mapping the photon number parity of the logical state to the ancilla state by applying a microwave pulse with multi-frequency components on the ancilla. Two components match the ancilla frequencies when the logical qubit is in the error space and other components are chosen symmetrically for the code space to eliminate the off-resonant driving effect on the logical states.
	\textbf{b} Bar chart of the measured photon number parities for the six cardinal-point states on the Bloch spheres of the logical qubit in the code and error spaces with the frequency comb parity measurement. Solid black frames correspond to the ideal parities $\pm1$ for the logical states in the code and error spaces, respectively. The numbers represent the average parity detection errors in these two spaces.
	\textbf{c} Measured Wigner function of the cavity state after encoding the logical qubit in the $\left\vert +X_\mathrm{L}\right\rangle$ state. 
	\textbf{d} and \textbf{e} Measured Wigner functions of the same cavity state after a waiting time of about 90~$\mu$s without (\textbf{d}) and with (\textbf{e}) a single QEC operation. The numbers in these Wigner functions represent the corresponding state fidelities.}
    \label{fig2}
\end{figure}

The key ingredients of a QEC procedure include encoding the quantum information to the logical qubit from the ancilla, the error syndrome measurement, the real-time error correction of the system depending on the measurement output, and the decoding process to readout the quantum information stored in the logical qubit. Our logical qubit is realized in a three-dimensional microwave cavity, and the dominant decoherence to combat is the excitation loss error. The logical qubit is encoded with a binomial code~\cite{michael2016}, with the codewords
\begin{eqnarray*}
\left\vert 0_\mathrm{L}\right\rangle  &=&\left( \left\vert 0\right\rangle +\left\vert 4\right\rangle \right) /\sqrt{2}\text{,} \\
\left\vert 1_\mathrm{L}\right\rangle  &=&\left\vert 2\right\rangle,
\end{eqnarray*}%
where the number in each ket denotes the photon number in the cavity. The binomial code is a typical stabilizer QEC code: When the single-photon-loss error occurs, the quantum information is projected into the error space spanned by \{$\left\vert 0_\mathrm{E}\right\rangle = \left\vert 3\right\rangle, \left\vert 1_\mathrm{E}\right\rangle = \left\vert 1\right\rangle $\}, with the photon number parity acting as the error syndrome to distinguish these two spaces. A general QEC protection of quantum information stored in the bosonic system is illustrated in Fig.~\ref{fig1}. After correctly measuring the photon number parity and applying the corresponding correction operations in real time, quantum information stored in the cavity can be recovered. 

The experiments are performed with a circuit QED architecture~\cite{blais2021}, where a superconducting transmon qubit~\cite{koch2007} as an ancilla is dispersively coupled to a three-dimensional microwave cavity~\cite{paik2011, axline2016, reagor2016}. The ancilla qubit has an energy relaxation time of about 98~$\mu$s and a pure dephasing time of 968~$\mu$s, while the storage cavity has a single-photon lifetime of 578~$\mu$s (corresponding to $\kappa_s/2\pi = 0.28$~kHz), and a pure dephasing time of 4.4~ms. The universal control of the multiple photon states of the cavity can be realized by utilizing the anharmonicity of the ancilla, and thus the key ingredients of the QEC procedure, as illustrated in Fig.~\ref{fig1}, to the logical qubit encoded in the high-dimensional Fock spaces of the bosnoic mode can be realized. 

Our route towards the break-even points in the QEC is two folds: improving both the operation fidelity to the logical qubit and the error syndrome measurement fidelity. The first goal is achieved by employing a tantalum transmon qubit with high coherence~\cite{place2021, wang2022} and an optimal quantum control technique~\cite{khaneja2005} with carefully calibrated system parameters [see Methods]. We attempt the second goal by an ingenious scheme of projection measurement of a selected collection of Fock states. The principle of the scheme is illustrated in Fig.~\ref{fig2}a, where a classical microwave pulse containing $2M$ frequency components is applied on the ancilla to read out the Fock states. Since the frequency of the ancilla is entangled with the photon number $n$ [see Methods for more details], error syndrome detection is achieved by mapping the even (odd) parity to the ancilla ground state $\left\vert g\right\rangle$ (excited state $\left\vert e\right\rangle$) in a quantum non-demolition manner. This approach holds potential advantages of more flexible choices of error spaces and less sensitive to ancilla damping and dephasing errors since the ancilla excitation is pronounced only when loss error happened. 

To characterize our syndrome measurement, the cavity is encoded to the six cardinal-point states in the Bloch spheres of both the code and error spaces based on the lowest-order binomial codewords. The measured results of the cavity photon number parities are displayed in Fig.~\ref{fig2}b and show an average detection error of 1.1\% and 2.5\% for the cavity states in the code and error spaces, respectively. The encoding of the cavity, one of the most elementary processes of QEC, is further verified by the Wigner function with a high fidelity of 0.95, as shown in Fig.~\ref{fig2}c.

Based on the above techniques, the QEC process of the binomial code can be implemented following the procedure in Fig.~\ref{fig1}. However, practical imperfections limit the QEC performance: (i) during a waiting time of $t_w$, i.e. an idle process, there is a probability of about $2(\kappa_s t_w)^2 \exp{(-2\kappa_s t_w)}$ for a two-photon-loss error, which is undetectable for this lowest-order binomial code. (ii) Due to the non-commutativity of the single-photon-loss error and the self-Kerr interaction of the cavity, there is a large dephasing effect of the logical qubit induced by the unpredictable photon loss event, thus destroying the stored quantum information. (iii) Quantum recovery operations are imperfect. It is worth noting that there is a logical state distortion even if no photon loss is detected~\cite{michael2016}. Considering the whole system, strategies to mitigate the above imperfections are introduced: choose an optimal waiting time, employ a two-layer QEC procedure~\cite{hu2019} to avoid unnecessary operation errors introduced by the error corrections, and adopt the photon-number-resolved a.c. Stark shift (PASS) method~\cite{ma2020} during idle operations to suppress the photon-jump-error-induced decoherence in the code space (see Supplementary Information for more details). The measured Wigner functions of the cavity states after a single QEC cycle (about 90 $\mu$s of waiting) without and with performing the error correction operation are shown in Fig.~\ref{fig2}(d, e), with state fidelities of 0.81 and 0.88 respectively.

\begin{figure*}
    \includegraphics{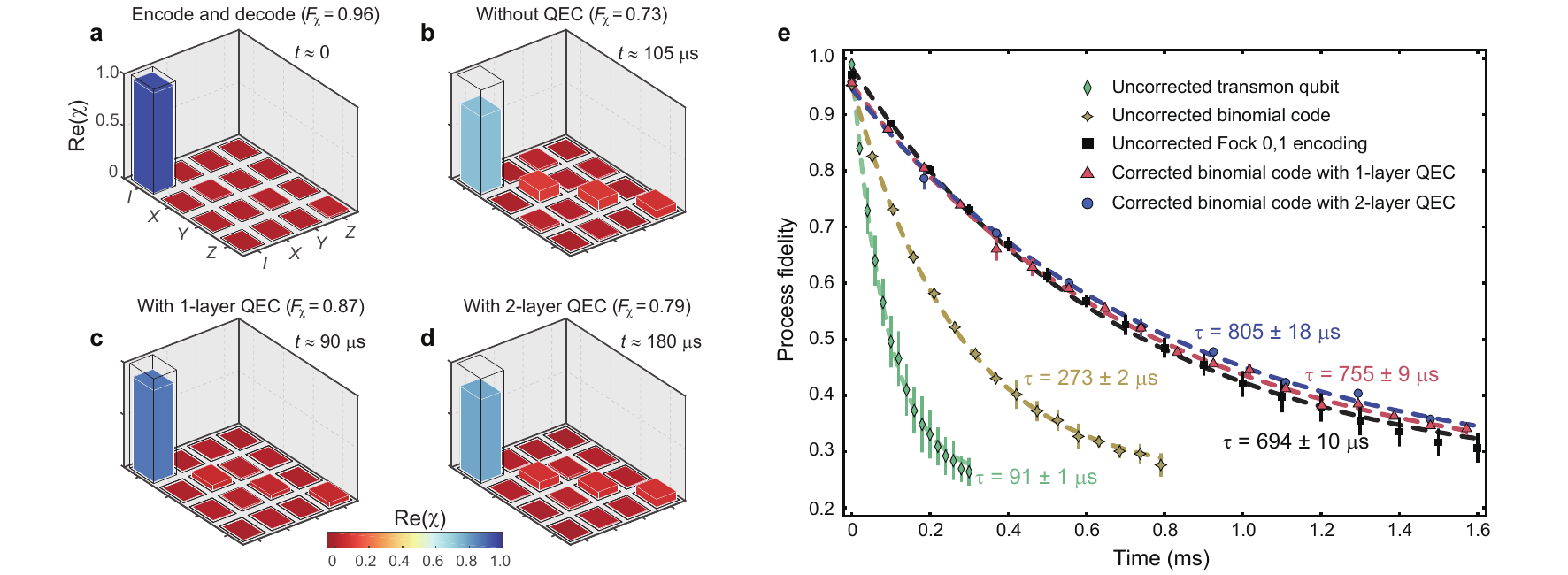}
    \caption{\textbf{Performance of repetitive QEC operations.}
	\textbf{a-d} Bar charts of the real parts of the process matrices for an encode and decode process~(\textbf{a}), a waiting of about $105~\mu$s without QEC (\textbf{b}), a cycle time of about $90~\mu$s with one-layer QEC operation (\textbf{c}), and a cycle time of about $180~\mu$s with two-layer QEC operation (\textbf{d}), respectively. Numbers in the brackets represent the process fidelities for each case.
	\textbf{e} Process fidelity decays as a function of time for different encodings. The process fidelities for both the corrected binomial code with one-layer QEC (red triangles) and two-layer QEC (blue circles) exhibit slow decay as compared to the uncorrected Fock states \{$\left\vert 0\right\rangle$, $\left\vert 1\right\rangle$\} encoding (black squares), which defines the break-even point in this system. The corrected binomial code with two-layer QECs offers an improvement over the break-even point by a factor of 1.2, and also surpasses the uncorrected binomial code (yellow stars) by a factor of 2.9 and the uncorrected transmon qubit (green diamonds) by a factor of 8.8. All curves are fitted using $F_\chi = Ae^{-t/\tau} + 0.25$ to extract the lifetimes $\tau$ of the corresponding encodings. Uncertainties on $\tau$ are obtained from the fittings.}
    \label{fig3}
\end{figure*}

The performance of the QEC is benchmarked by the process fidelity $F_\chi = \mathrm{tr}\left( \chi_\mathrm{exp} \chi_\mathrm{ideal} \right)$, which is defined by comparing the experimental measured process matrix $\chi_\mathrm{exp}$ for the QEC process with the ideal process matrix $\chi_\mathrm{ideal}$ for an identity operation. In Fig.~\ref{fig3}a, we present the measured process matrix for the encoding and decoding process only, which indicates a reference fidelity of 0.96. In the absence of a QEC operation after a waiting time of 105 $\mu$s, the process fidelity is reduced to a value of 0.73 due to the inability to protect the quantum information stored in the cavity from the single-photon-loss error, with the corresponding measured process matrix shown in Fig.~\ref{fig3}b. When utilizing the QEC operation, the process fidelity is indeed improved due to the protection from the single-photon-loss error, with the process matrices for the one- and two-layer QECs shown in Fig.~\ref{fig3}(c, d), respectively. 

The most important benchmark to characterize the performance of a QEC procedure is the gain in the lifetime of the protected logical qubit against that of the constituent element with the longest lifetime. For the 3D circuit QED device, the best physical qubit is encoded with the two lowest photon-number states \{$\left\vert 0\right\rangle $, $\left\vert 1\right\rangle $\}, which is more robust against decoherence effects than any other encoded photonic qubit without QEC protection. To quantitatively show the advantage of our QEC scheme, in Fig.~\ref{fig3}e we display the measured process fidelities of the corrected binomial code as a function of the storage time with the repetitive one-layer (red triangles) and two-layer (blue circles) QECs, as well as those for the unprotected binomial code (yellow stars), the transmon qubit (green diamonds), and the Fock qubit (black squares) for comparison. 

All curves are fitted according to the function $F_\chi = Ae^{-t/\tau} + 0.25$ with $\tau$ corresponding to the lifetime of the specific encoding and $A$ being a fitting parameter. The offset in the fitting function is fixed to 0.25, implying a complete loss of information at the final time. As a result, the lifetime $\tau$ for the corrected binomial code with one-layer QECs is improved by about 8.3 times compared to the uncorrected transmon qubit, 2.8 times compared to the uncorrected binomial code. In particular, $\tau$ is about $0.1$ times longer than the uncorrected Fock qubit encoding, i.e., exceeds the break-even point of QEC in this system. Employing the two-layer QEC scheme, the corresponding lifetime $\tau$ of the logical qubit is improved to about 8.8 times that of the uncorrected transmon qubit, 2.9 times that of the uncorrected binomial code, and 1.2 times that of the break-even point. These results demonstrate that the quantum information stored in the cavity with multiphoton binomial encoding can indeed be preserved and protected from the photon loss errors via repetitive QEC operations. 

Table~\ref{error_budget} shows an overall error analysis for the one- and two-layer QEC experiments. The error sources are divided into four parts: the intrinsic errors for the lowest-order binomial code, the error detection errors, the recovery operation errors, and the ancilla thermal excitation errors during the QEC cycle. These errors can be estimated from either the numerical simulations, or the measurement results of individual calibration experiments (see Supplementary Information). The predicted lifetimes $\tau$ for the QEC experiments, calculated by $\tau = -T_w/\ln(1-\epsilon)$~\cite{hu2019}, with $T_w$ and $\epsilon$ being the total duration and the weighted total error per QEC cycle, are well consistent with those in our QEC experiments.

\begin{table}[tb]
\caption{Error budget for the one- and two-layer QEC processes. The predicted lifetime calculated from the error model agrees well with the measured lifetimes. *These errors are estimated from numerical simulations.}
\centering
\begin{tabular}{c|cc}
  \hline
  \hline
  Error source &  1-layer QEC & 2-layer QEC \\\hline
  &&\\[-0.9em]
  Intrinsic error* &   6.4\% &  12.4\%  \\
  &&\\[-0.9em]
  Detection error &  1.4\% & 2.8\%  \\
  &&\\[-0.9em]
  Recovery error &  2.9\% & 3.8\%  \\
  &&\\[-0.9em]
  Thermal error &  0.8\% & 1.1\%  \\
  &&\\[-0.9em]
  Total error &  11.5\% & 20.1\%  \\
  &&\\[-0.9em]
  \hline
  &&\\[-0.9em]
  Predicted lifetime &  757~$\mu$s & 824~$\mu$s  \\
  &&\\[-0.9em]
  Measured lifetime &  $755\pm 9$~$\mu$s & $805\pm 18$~$\mu$s  \\
  &&\\[-0.9em]
  \hline 
  \hline
\end{tabular}
\label{error_budget}
\end{table}

In conclusion, we experimentally demonstrate the prolonged coherence time of quantum information encoded with discrete variables in a bosonic mode by repetitive QEC. The break-even point has been reached by carefully designing the QEC procedure to balance the fidelity losses due to the undetectable errors during the idle process and the error detection and correction operations. 
Presently, the main infidelity is contributed by the two-photon-loss error that is beyond the ability of our current QEC code, but can be corrected by higher-order binomial codes~\cite{michael2016}. Our frequency comb method could naturally be utilized to measure the generalized photon number parity of such codes, enabling detection and correction of both single- and two-photon-loss errors. 
Our work thus represents a key step towards scalable quantum computing and provides a practical guide for system optimization of quantum control and the design of the QEC procedure for future applications of logical qubits.

\textit{Note added}--After completion of the experiment, we became aware of a QEC experiment, which also goes beyond the break-even point, but is based on a continuous-variable encoding scheme.


\smallskip{}

\noindent \textbf{\large{}Online content}{\large\par}

\noindent Any methods, additional references, Nature Research reporting
summaries, source data, extended data, supplementary information,
acknowledgements, peer review information; details of author contributions
and competing interests; and statements of data and code availability
are available on line.

%

\clearpage{}
\setcounter{figure}{0} 
\noindent \textbf{\large{}Methods}{\large\par}

\noindent \textbf{Experimental device and setup}

\noindent The circuit QED device in our experiment uses a hybrid 3D-planar architecture~\cite{axline2016} and consists of a superconducting transmon qubit~\cite{koch2007}, a coaxial stub cavity, and a Purcell-filtered stripline readout resonator (Fig.~S1 in Supplementary Information section I). The high-Q cavity is designed with a cylindrical re-entrant quarter-wave transmission line resonator~\cite{reagor2016}, and machined from high purity (99.9995\%) aluminum. A horizontal tunnel is used to house a sapphire chip, on which the antenna pads of the transmon qubit and the striplines of the low-Q readout resonator are patterned with a thin-film tantalum~\cite{place2021, wang2022}. The single Al-AlO$_x$-Al trilayer Josephson junction of the transmon qubit is fabricated using a double angle evaporation technique. 

The fast feedback control is implemented with Zurich Instruments UHFQA and HDAWG, which are connected to each other through a DIO link cable for real time feedback control. The UHFQA generates the readout pulses, acquires the down-converted transmitted readout signals for demodulation and discrimination in hardware, and sends the digitized readout results to the HDAWG through the DIO link cable in real time. The HDAWG plays different predefined waveforms conditional on the received readout results from the DIO link cable. The feedback latency, defined as the time interval between sending out the last point of the readout pulse from UHFQA and sending out the first point of the control pulse from HDAWG, is about 511~ns in our setup, which also includes the signal travelling time through the experimental circuitry.

\vbox{}

\noindent \textbf{Parity mapping} 

\noindent 
The parity mapping procedure in the QEC experiment is implemented by applying a classical microwave pulse containing $2M$ ($M=11$ in our experiment) frequency components on the ancilla qubit, with the system dynamics governed by the Hamiltonian  
\begin{equation}
H/\hbar=- \chi a^{\dagger }a\left\vert e\right\rangle \left\langle
e\right\vert + \Omega \left[ \sum_{n=1}^{2M}e^{-i\delta _{n}t}\left\vert e\right\rangle \left\langle g\right\vert +\mathrm{h.c.}\right] ,
\end{equation}
in the interaction picture. Here, $\left\vert e\right\rangle $ ($\left\vert g\right\rangle $) denotes the excited (ground) state of the ancilla qubit, $a^{\dagger }$ ($a$) is the creation (annihilation) operator of the photonic field in the cavity, $\chi $ is the ancilla's frequency shift induced per photon due to their dispersive coupling, $\delta_{n}$ is the frequency detuning of the $n$-th driving component with a Rabi frequency of $\Omega $, and $\mathrm{h.c.}$ denotes the Hermitian conjugate. With the choice of the drive frequency detuning $\delta_n = (2M-2n-1)\chi$, the ancilla is resonantly driven when the cavity have $2m+1$ photons with $m=0,1,...M$.

For the cavity in the code space, the ancilla is off-resonantly driven by the comb pulse. For the 2-photon state in the cavity, the qubit's transition $\left\vert g\right\rangle \leftrightarrow \left\vert e\right\rangle $ is driven by $M$ pairs of frequency components with symmetric detunings, resulting in a qubit state revival at a time of $T = k\pi/\chi$ with $k$ being an integer. Similarly, for the 0- and 4-photon states in the cavity, the qubit is driven by $M-1$ pairs of symmetric components and two unpaired components, whose effects can be ignored under the condition of $2M\chi\gg \Omega$. Therefore, the ancilla also makes a cyclic evolution at $T = k\pi/\chi$ and returns to the initial ground state when the cavity is in the code space.  

For the cavity in the error space with 1- and 3-photon states, the ancilla qubit's transition $\left\vert g\right\rangle \leftrightarrow \left\vert e\right\rangle $ is driven by a resonant frequency component, $M-1$ pairs of symmetric frequency components, and an unpaired off-resonant component. Under the same condition of $2M\chi\gg \Omega$, we can neglect the off-resonant effect of the unpaired components, and the ancilla will evolve from the initial ground state to the excited state at $T = k\pi/\chi$, with $k$ being an integer when choosing the drive amplitude $\Omega = \pi/2T$. In our experiment, $\Omega = \chi/4$, while $T\approx \pi/\chi$ for an optimized parity mapping time (see Supplementary Informations). 

Therefore, this frequency comb pulse achieves error syndrome detection by mapping the even (odd) parity of the cavity state to the ancilla $\left\vert g\right\rangle$ ($\left\vert e\right\rangle$) state in a QND manner. This parity mapping process can be intuitively illustrated by simultaneously applying two conditional $\pi$ rotations to the ancilla qubit to flip the qubit state to the excited state associated with the cavity's 1- and 3-photon states, thus resulting in a minimum perturbation to the cavity states in the code space. 

\vbox{}
\noindent \textbf{Strategies for system optimization}

\noindent 
The photon-number-resolved a.c. Stark shift method~\cite{ma2020} is adopted to mitigate the photon-loss-induced dephasing effect of the logical codewords, due to the non-commutativity of the annihilation operation and the self-Kerr term. In our experiment, we apply an off-resonant drive pulse with a frequency detuning of about $-3.5\chi$ on the ancilla during the idle operation, resulting in different phase accumulation rate $f_n$ for Fock state $\left| n \right \rangle$ with $n=1,2,3,4$ relative to the vacuum state. By choosing an optimal amplitude of the detuned drive, we could achieve the error-transparent condition~\cite{ma2020} of $(f_4-f_2) - (f_3 -f_1) = 0$ to mitigate the dephasing effect of the logical qubit (Fig.~S4 in the Supplementary Information of Section III).

In order to balance the operation errors, no-parity-jump backaction errors, and photon-loss errors, we employ a two-layer QEC procedure~\cite{hu2019} to improve the QEC performance (Fig.~S6 in the Supplementary Information of Section III). In our QEC experiment, there are two bottom layers in a single QEC cycle, where the first one conserves the photon number parity in the deformed code space, and the second one recovers the quantum information in the code space.

The waiting time of the idle operation in each QEC cycle is selected based on a tradeoff between the uncorrected errors occurring during this time and the operation errors occurring during the error syndrome measurements and recovery operations. On one hand, the longer the waiting time, the larger probability of the two-photon-loss event is during this time, which cannot be detected by the lowest-order binomial code. On the other hand, the more frequent the error detection, the more likely the photon-loss errors occur during the detections and corrections. We calculate the QEC lifetime as a function of the waiting time from numerical simulations and choose an optimal waiting time of about $90~\mu$s in our QEC experiment. (Fig.~S8 in the Supplementary Information of Section IV).

\vbox{}
\smallskip{}

\noindent \textbf{\large{}Data availability}{\large\par}

\noindent All data generated or analysed during this study are available
within the paper and its Supplementary Information. Further source
data will be made available on reasonable request.

\smallskip{}

\noindent \textbf{\large{}Code availability}{\large\par}

\noindent The code used to solve the equations presented in the Supplementary
Information will be made available on reasonable request.

\smallskip{}

\noindent \textbf{\large{}Acknowledgment}{\large\par}

\noindent This work was supported by the Key-Area Research and Development Program of Guangdong Province (Grants No. 2018B030326001 and No. 2020B0303030001), the Shenzhen Science and Technology Program (Grant No. RCYX20210706092103021), the National Natural Science Foundation of China (Grants No. 12274198, No. 11904158, No. U1801661, No. 12274080, No. 12061131011, No. 92265210, No. 92165209, No. 11925404, No. 11890704 and No. 11875108), the Guangdong Basic and Applied Basic Research Foundation (Grant No. 2022A1515010324), the Guangdong Provincial Key Laboratory (Grant No. 2019B121203002), the Program (Grant No. 2016ZT06D348), the Science, Technology and Innovation Commission of Shenzhen Municipality (Grant No. KYTDPT20181011104202253), the Shenzhen-Hong Kong cooperation zone for technology and innovation (Contract No. HZQB-KCZYB-2020050), the National Key Research and Development Program of China (Grant No. 2017YFA0304303), the China Postdoctoral Science Foundation (BX2021167), the Innovation Program for Quantum Science and Technology (Grants No. ZD0301703 and No. ZD0102040201), and the Natural Science Foundation of Beijing (Grant No. Z190012). 

\smallskip{}

\noindent \textbf{\large{}Author contributions}{\large\par}

\noindent Y.X. and D.Y. supervised the project. Y.X. conceived and designed the experiment. Z.N. performed the experiment. Z.N. and Y.X. analyzed the data and carried out the numerical simulations. Z.N. and Sa.L. developed the feedback control technique under the supervision of Y.X. X.D., Y.C., W.W., Z.-B.Y., and F.Y. contributed to the experimental and theoretical optimization. L.Z., So.L. and H.Y. provided supports in device fabrication. S.-B.Z. proposed the theoretical scheme of the frequency comb method. S.-B.Z., C.-L.Z. and L.S. provided theoretical and experimental supports. C.-L.Z., S.-B.Z., L.S., and Y.X. wrote the manuscript with feedback from all authors.

\smallskip{}

\noindent \textbf{\large{}Competing interests}{\large\par}

\noindent The authors declare no competing interests.

\smallskip{}

\noindent \textbf{\large{}Additional information}{\large\par}

\noindent \textbf{Supplementary information} The online version contains
supplementary material.

\noindent \textbf{Correspondence and requests for materials} should
be addressed to L.S., S.-B.Z., Y. X. or D.Y. 

\clearpage{}

\end{document}


\title{Supplementary Information for ``Beating the break-even point with a discrete-variable-encoded logical qubit"}

\author{Zhongchu Ni}
\affiliation{Shenzhen Institute for Quantum Science and Engineering, Southern University of Science and Technology, Shenzhen 518055, China}
\affiliation{International Quantum Academy, Shenzhen 518048, China}
\affiliation{Guangdong Provincial Key Laboratory of Quantum Science and Engineering, Southern University of Science and Technology, Shenzhen 518055, China}
\affiliation{Department of Physics, Southern University of Science and Technology, Shenzhen 518055, China}

\author{Sai Li}
\author{Xiaowei Deng}
\author{Yanyan Cai}
\author{Libo Zhang}
\affiliation{Shenzhen Institute for Quantum Science and Engineering, Southern University of Science and Technology, Shenzhen 518055, China}
\affiliation{International Quantum Academy, Shenzhen 518048, China}
\affiliation{Guangdong Provincial Key Laboratory of Quantum Science and Engineering, Southern University of Science and Technology, Shenzhen 518055, China}

\author{Weiting Wang}
\affiliation{Center for Quantum Information, Institute for Interdisciplinary Information Sciences, Tsinghua University, Beijing 100084, China}

\author{Zhen-Biao Yang}
\affiliation{Fujian Key Laboratory of Quantum Information and Quantum Optics, College of Physics and Information Engineering, Fuzhou University, Fuzhou, Fujian 350108, China}

\author{Haifeng Yu}
\affiliation{Beijing Academy of Quantum Information Sciences, Beijing 100193, China}

\author{Fei Yan}
\author{Song Liu}
\affiliation{Shenzhen Institute for Quantum Science and Engineering, Southern University of Science and Technology, Shenzhen 518055, China}
\affiliation{International Quantum Academy, Shenzhen 518048, China}
\affiliation{Guangdong Provincial Key Laboratory of Quantum Science and Engineering, Southern University of Science and Technology, Shenzhen 518055, China}

\author{Chang-Ling Zou}
\affiliation{CAS Key Laboratory of Quantum Information, University of Science and Technology of China, Hefei, Anhui 230026, China}
\author{Luyan Sun}
\email{luyansun@tsinghua.edu.cn}
\affiliation{Center for Quantum Information, Institute for Interdisciplinary Information Sciences, Tsinghua University, Beijing 100084, China}

\author{Shi-Biao Zheng}
\email{t96034@fzu.edu.cn}
\affiliation{Fujian Key Laboratory of Quantum Information and Quantum Optics, College of Physics and Information Engineering, Fuzhou University, Fuzhou, Fujian 350108, China}

\author{Yuan Xu}
\email{xuy5@sustech.edu.cn}
\affiliation{Shenzhen Institute for Quantum Science and Engineering, Southern University of Science and Technology, Shenzhen 518055, China}
\affiliation{International Quantum Academy, Shenzhen 518048, China}
\affiliation{Guangdong Provincial Key Laboratory of Quantum Science and Engineering, Southern University of Science and Technology, Shenzhen 518055, China}

\author{Dapeng Yu}
\email{yudp@sustech.edu.cn}
\affiliation{Shenzhen Institute for Quantum Science and Engineering, Southern University of Science and Technology, Shenzhen 518055, China}
\affiliation{International Quantum Academy, Shenzhen 518048, China}
\affiliation{Guangdong Provincial Key Laboratory of Quantum Science and Engineering, Southern University of Science and Technology, Shenzhen 518055, China}
\affiliation{Department of Physics, Southern University of Science and Technology, Shenzhen 518055, China}

\maketitle
\vskip 0.5cm

\section{Experimental method}
\subsection{Device}
\begin{figure*}[tb]
  \centering
  \includegraphics{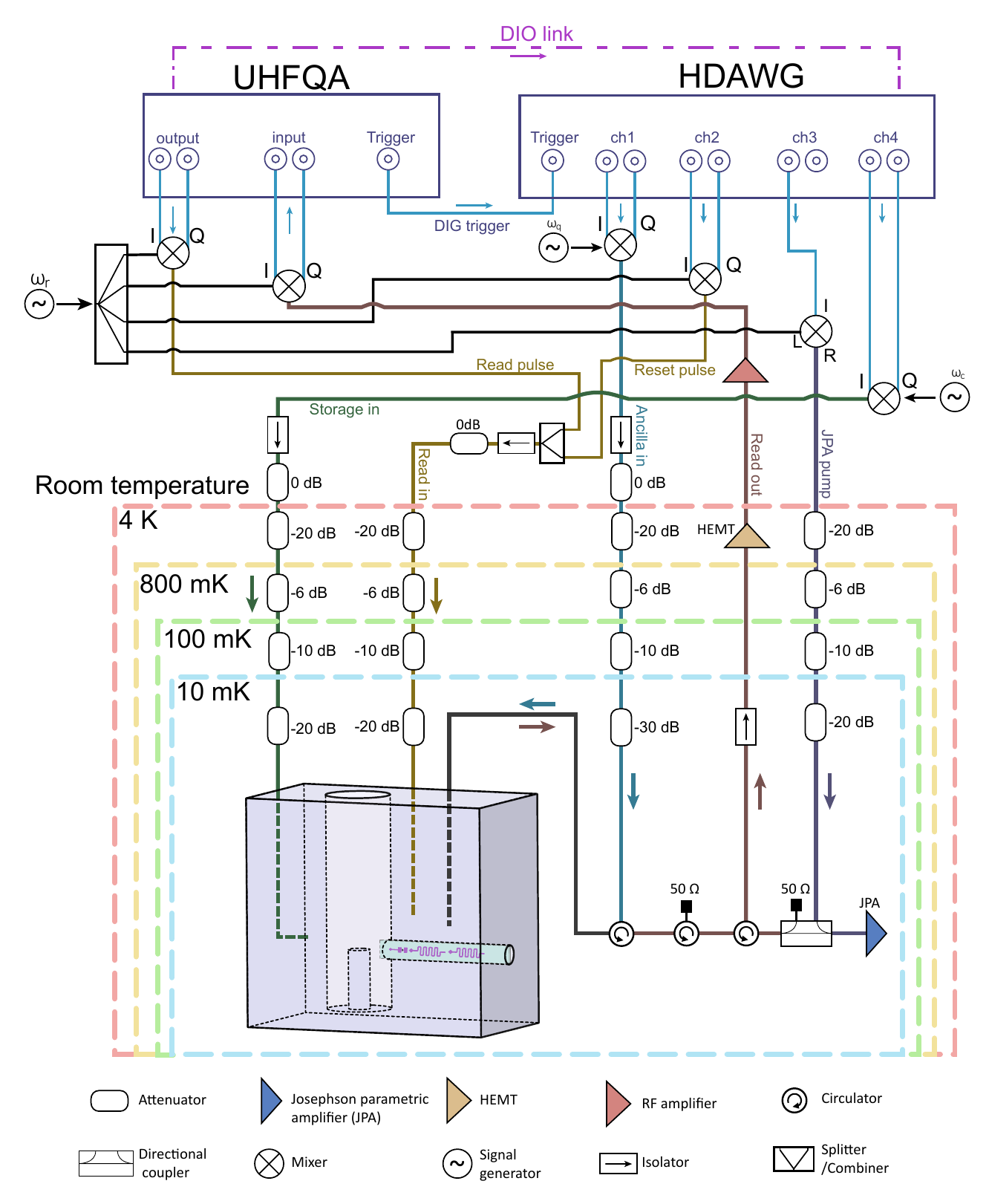}
  \caption{Full wiring of the experimental circuitry and device schematic.}
  \label{fig_setup}
\end{figure*}

The quantum error correction (QEC) experiment is implemented in a three-dimensional (3D) circuit quantum electrodynamics (QED) architecture~\cite{wallraff2004, blais2004, blais2021, paik2011,kirchmair2013,vlastakis2013}, which consists of a superconducting transmon qubit~\cite{koch2007}, a 3D coaxial stub cavity~\cite{reagor2016, wang2016, gao2019}, and a Purcell-filtered stripline readout resonator~\cite{axline2016, chou2018}. A schematic of the device is shown in Fig.~\ref{fig_setup}. The 3D circuit QED device is directly machined from a single block of high purity (5N5) aluminum and chemically etched to improve the cavity's coherence time~\cite{reagor2013}. 

The coaxial stub cavity is constructed as a 3D $\lambda/4$ transmission line resonator with the fundamental mode used for storing microwave photons and encoding the bosonic logical qubit, henceforth referred to as the storage cavity. The Purcell-filtered readout resonator is constructed with two quasi-planar $\lambda/2$ transmission line resonators, which are formed by the metal wall of a horizontal tunnel and two metal strips on the qubit chip inserted in the tunnel. The transmon qubit is patterned on a sapphire chip with two antenna pads to couple to the storage cavity mode and the stripline readout resonator mode. The Josephson junction of the qubit is an $\mathrm{Al}-\mathrm{Al}_2\mathrm{O}_3-\mathrm{Al}$ trilayer tunnel junction formed by a double angle evaporation technique, and the antenna pads and readout striplines are grown using tantalum films in BCC alpha-phase to improve the coherence time of the transmon qubit~\cite{place2021, wang2022}. 

In our experiment, the transmon qubit serves as an auxiliary qubit for error detection and correction operations of the bosonic logical qubit in the storage cavity. The stripline readout resonator is strongly coupled to the transmon qubit for fast dispersive readout of the qubit states, and coupled to the outside world via another quasi-planar $\lambda/2$ transmission line resonator, denoted as the Purcell filter resonator, to protect the coherence times of both the auxiliary qubit and the storage cavity. To fit both the readout resonator and filter resonator, the striplines are designed with wiggles to decrease the physical footprint of the patterns on the sapphire chip~\cite{chou2018}.

\subsection{Setup}
The experimental device is covered with a magnetic shield and installed inside a cryogenic-free dilution refrigerator, which has a temperature below 10~mK. Both the qubit and bosonic logical qubit in the storage cavity are controlled with microwave pulses generated by single-sideband in-phase and quadrature (IQ) modulations. The corresponding waveforms for each mode are generated from two digital-to-analog converter (DAC) channels of the Zurich Instruments high-density arbitrary waveform generator (ZI HDAWG). The qubit control pulses have a cosine-shaped envelope with a duration of about 20~ns, in combination with the technique of ``derivative removal by adiabatic gate" (DRAG) to remove leakage errors to higher energy levels~\cite{motzoi2009, gambetta2011}. The cavity control pulses for encoding and decoding the bosonic logical qubit and implementing the recovery operations after each error detection are generated from numerical optimization with the gradient ascent pulse engineering (GRAPE) method~\cite{khaneja2005}. The readout pulse is generated by single-sideband IQ modulation of the waveforms from two DAC channels of a Zurich Instruments ultra-high frequency quantum analyzer (ZI UHFQA) in cooperation with a signal generator as the local oscillator and sent to the readout resonator through a coaxial cable with a series of attenuators and filters. 

The transmitted signal from the resonator is amplified by a series amplification chain, including a quantum limited Josephson parametric amplifier (JPA) at the base temperature, a high electron mobility transistor (HEMT) at the 4K stage, and a standard commercial low-noise RF amplifier at room temperature. Finally, the signal is downconverted by an IQ mixer with the same local oscillator as that used to generate the readout pulse. The downconverted IQ signals are digitized and recorded by the analogy-to-digital converter (ADC) of the same ZI UHFQA. To align the timing sequence of the qubit and cavity control pulses with the readout pulse, the UHFQA serves as a master instrument to send a trigger signal to control the slave instrument HDAWG. Figure~\ref{fig_setup} shows a schematic of the full wiring of the experimental setup. 

The QEC experiment requires both error syndrome measurements and error corrections in real time, which is realized by connecting the UHFQA to the HDAWG with a digital input/output (DIO) link cable. The UHFQA has the capabilities of performing the demodulation and calculations of the downconverted readout signals in hardware and discriminating the results into digitized signals in real time. These digital signals are not only sent to the host PC, but also to the HDAWG for fast feedback control through the DIO link. The HDAWG can store a series of predefined waveforms and selectively play waveforms conditional on the received DIO signal in real time, which realizes the quantum feedback control for quantum error correction operations with a minimized latency. The duration of sending the last point of the readout pulse from UHFQA and sending out the first point of the feedback control signal is defined as the feedback latency, which is about 511~ns in our experiment (including the signal travelling time through the experimental circuitry).

\begin{table}[tb]
\caption{Hamiltonian parameters and coherence times.}
\centering
\begin{tabular}{p{3cm}<{\centering} p{3cm}<{\centering}  p{3cm}<{\centering} p{3cm}<{\centering} p{3cm}<{\centering}}
  \hline
  \hline
  \multicolumn{2}{c}{Measured (predicted) parameters} & Transmon qubit & Storage cavity & Readout resonator \\\hline
  &&&\\[-0.9em]
  \multicolumn{2}{c}{Mode frequency $\omega_\mathrm{q,c,r}/2\pi$ } & 4.962~GHz & 6.532~GHz & 8.562~GHz  \\
  &&&\\[-0.9em]
  \hline
  &&&\\[-0.9em]
  \multirow{3}{*}{Kerr interactions} & Transmon qubit & 216~MHz & 2.59~MHz & 1.9~MHz  \\
                                     & Storage cavity & 2.59~MHz & 9.7~kHz & (12.7~kHz) \\
                                     & Readout resonator & 1.9~MHz & (12.7~kHz) & (4.2~kHz) \\
  &&&\\[-0.9em]
  \hline
  &&&\\[-0.9em]
  \multicolumn{2}{c}{Higher-order self-Kerr $K'_\mathrm{c}/2\pi$ } & & 0.32~kHz &  \\
  &&&\\[-0.9em]
  \multicolumn{2}{c}{Higher-order cross-Kerr $\chi'_\mathrm{qc}/2\pi$ } & \multicolumn{2}{c}{5.41~kHz} &  \\
  &&&\\[-0.9em]
  \hline
  &&&\\[-0.9em]
  \multicolumn{2}{c}{Relaxation time $T_1$ } &  98~$\mu$s~ & 578~$\mu$s  & 58~ns  \\
  &&\\[-0.9em]
  \multicolumn{2}{c}{Pure dephasing time $T_\phi$ } & 968~$\mu$s & 4389~$\mu$s & -   \\
  &&\\[-0.9em]
  \multicolumn{2}{c}{Thermal population} & 1.3\% & 0.6\% & $<0.1\%$ \\
  &&\\[-0.9em]
  \hline 
  \hline
\end{tabular}
\label{TableS1}
\end{table}

\subsection{System Hamiltonian}
The Hamiltonian of our system containing a storage cavity, a readout resonator, and an auxiliary qubit can be expressed in the dispersive regime as
\begin{eqnarray}
\label{H1}
\hat{H}/\hbar &=& \omega_\mathrm{q} \hat{a}_\mathrm{q}^\dag \hat{a}_\mathrm{q} + \omega_\mathrm{c} \hat{a}_\mathrm{c}^\dag \hat{a}_\mathrm{c} +\omega_\mathrm{r} \hat{a}_\mathrm{r}^\dag \hat{a}_\mathrm{r} \notag\\
&-& \frac{K_\mathrm{q}}{2}\hat{a}_\mathrm{q}^{\dag 2}\hat{a}_\mathrm{q}^{2}-\frac{K_\mathrm{c}}{2}\hat{a}_\mathrm{c}^{\dag 2}\hat{a}_\mathrm{c}^{2} + \frac{K'_\mathrm{c}}{6}\hat{a}_\mathrm{c}^{\dag 3}\hat{a}_\mathrm{c}^{3}-\frac{K_\mathrm{r}}{2}\hat{a}_\mathrm{r}^{\dag 2}\hat{a}_\mathrm{r}^{2}\notag\\
&-& \chi_\mathrm{qr}\hat{a}_\mathrm{q}^\dag\hat{a}_\mathrm{q}\hat{a}_\mathrm{r}^\dag\hat{a}_\mathrm{r} -\chi_\mathrm{cr}\hat{a}_\mathrm{c}^\dag\hat{a}_\mathrm{c}\hat{a}_\mathrm{r}^\dag\hat{a}_\mathrm{r} - \chi_\mathrm{qc}\hat{a}_\mathrm{q}^\dag\hat{a}_\mathrm{q}\hat{a}_\mathrm{c}^\dag\hat{a}_\mathrm{c} + \frac{\chi'_\mathrm{qc}}{2}\hat{a}_\mathrm{q}^\dag\hat{a}_\mathrm{q}\hat{a}_\mathrm{c}^{\dag 2}\hat{a}_\mathrm{c}^{2}
,
\end{eqnarray}
where $\omega_\mathrm{q,c,r}$ are the frequencies of the auxiliary qubit, storage cavity, and readout resonator, respectively; $\hat{a}_\mathrm{q,c,r}$($\hat{a}^\dag_\mathrm{q,c,r}$) are their corresponding annihilation (creation) operators; $K_\mathrm{q,c,s}$ are the self-Kerrs of the corresponding mode; $K'_{c}$ is a higher-order self-Kerr of the storage cavity; $\chi_\mathrm{qr}$, $\chi_\mathrm{cr}$, and $\chi_\mathrm{qc}$ are the cross-Kerrs between these three modes; and  $\chi'_\mathrm{qc}$ is the higher-order cross-Kerr between the auxiliary qubit and the storage cavity. All these parameters are summarized and listed in Table~\ref{TableS1}. Note that some parameters cannot be directly measured and are predicted according to $\chi_{ab} = 2\sqrt{K_a K_b}$ based on  black-box quantization (BBQ) theory~\cite{nigg2012}.

When only considering the transmon qubit in the two lowest energy levels and discarding the readout resonator mode during the parity mapping process in the QEC experiment, the above Hamiltonian can be rewritten as
\begin{eqnarray}
\label{H2}
\hat{H}_\mathrm{qc}/\hbar = (\omega_\mathrm{q}-\chi_\mathrm{qc}\hat{a}_\mathrm{c}^\dag\hat{a}_\mathrm{c})\ket{e}\bra{e}+\omega_\mathrm{c} \hat{a}_\mathrm{c}^\dag \hat{a}_\mathrm{c}-\frac{K_\mathrm{c}}{2}\hat{a}_\mathrm{c}^{\dag 2}\hat{a}_\mathrm{c}^{2} , 
\end{eqnarray}
where $\ket{e}$ ($\ket{g}$) denotes the auxiliary qubit excited (ground) state. With this dispersive interaction between the auxiliary qubit and the storage cavity, the qubit transition frequency is modified by the photon number in the cavity with an energy level spacing of $\chi_{qc}$ per photon. This homogeneous feature allows for implementing the parity mapping procedure with the frequency comb control method, which will be discussed in detail in Sec.~\ref{theory}.

\subsection{Coherence times and readout fidelities}
The coherence times and thermal populations of the qubit and cavity modes are experimentally measured and are also summarized in Table~\ref{TableS1}. The auxiliary qubit has an average energy relaxation time of $T_1 = 98~\mu$s and a Ramsey coherence time of $T^*_2 = 163~\mu$s, inferring a pure dephasing time of $T_\phi = 968~\mu$s, while the storage cavity has a single-photon lifetime of $T_1 = 578~\mu$s ($\kappa_c/2\pi = 0.28$~kHz) and a Ramsey coherence time of $T^*_2 = 915~\mu$s, corresponding to a pure dephasing time of $T_\phi = 4389~\mu$s. The readout resonator is designed to have an energy relaxation time of about 58~ns ($\kappa_r/2\pi = 2.7$~MHz) for fast single-shot qubit readout. The pure dephasing time of the storage cavity is about seven times larger than the single-photon energy relaxation time, indicating that the photon-loss error is the dominant error source for the logical qubit. The thermal populations of the auxiliary qubit and the storage cavity are $n^\mathrm{q}_\mathrm{th} = 1.3\%$ and $n^\mathrm{c}_\mathrm{th} = 0.6\%$, respectively. The thermal population of the readout resonator $n^\mathrm{r}_\mathrm{th}$ is inferred by the qubit pure dephasing time $T_\phi < 1/n^\mathrm{r}_\mathrm{th} \kappa_r$~\cite{sears2012}, giving an upper bound of the readout resonator thermal population of $n^\mathrm{r}_\mathrm{th}<0.1\%$. 

The dispersive interaction between the auxiliary qubit and the readout resonator allows for quantum non-demolition (QND) measurement of the qubit states, which is achieved by measuring the transmission signals through the readout resonator. The decay rate of the readout resonator $\kappa_r/2\pi = 2.7$~MHz is designed to match the dispersive coupling strength $\chi_\mathrm{qr}/2\pi = 1.9$~MHz between the qubit and readout resonator to improve the signal-to-noise ratio of the readout signals and achieve the high-fidelity (with the help of a JPA) and high-QND single-shot readout of the auxiliary qubit. The readout pulse has an optimized duration of about 600~ns, giving an average readout fidelity of 0.993 (0.998 for the $|g\rangle$ state and 0.988 for the $|e\rangle$ state), and an average QNDness of 0.985 (0.998 for the $|g\rangle$ state and 0.972 for the $|e\rangle$ state). 

\subsection{Quantum optimal control}
The control pulses of the storage cavity for encoding and decoding the bosonic logical qubit and implementing the recovery operations are generated from quantum optimal control with GRAPE technique~\cite{khaneja2005}. Each ideal unitary operation $U_\mathrm{ideal}$ is implemented by applying two control pulses $\epsilon_q(t)$ and $\epsilon_c(t)$ to drive both the auxiliary qubit and the cavity, in order to realize a set of simultaneous state transfers to map each initial state $| \psi_0^i\rangle$ to the corresponding final state $| \psi_f^i\rangle = U_\mathrm{ideal}| \psi_0^i\rangle$ for the $i$-th initial state in the relevant subspace. The goal for the quantum control is to maximize the average fidelity $F = \left|\sum_i{\langle \psi_f^i | U(\epsilon_q(t), \epsilon_c(t)) | \psi_0^i \rangle} \right|^2$ of these state transfers by optimizing the control pulses $\epsilon_q(t)$ and $\epsilon_c(t)$. Here, $U(\epsilon_q(t), \epsilon_c(t))$ is the unitary evolution operator with the corresponding drive pulses applied on the system.

In order to numerically solve this optimization problem, we assume these control pulses are piecewise constant functions by dividing the total gate time $T$ into $N$ segments with each duration of $\Delta t = T/N$. Then the total evolution operator $U$ can be expressed as $U(\epsilon_q(t), \epsilon_c(t)) = U_N U_{N-1}...U_2U_1$, where $U_k = \exp{\left(-i(H_0 + H_d)\Delta t/\hbar\right)}$ represents the evolution operator of the $k$-th segment. Here $H_0$ is the drift Hamiltonian, describing the dispersive interaction between the auxiliary qubit and the storage cavity, $H_d = \epsilon^I_q(k\Delta t) \sigma_x + \epsilon^Q_q(k\Delta t) \sigma_y + \epsilon^I_c(k\Delta t) (\hat{a}_\mathrm{c} + \hat{a}_\mathrm{c}^\dag) + i\epsilon^Q_c(k\Delta t) (\hat{a}_\mathrm{c} - \hat{a}_\mathrm{c}^\dag)$ is the driving Hamiltonian for both the auxiliary qubit and the cavity at the $k$-th segment, and $\epsilon^I$ and $\epsilon^Q$ are the in-phase and quadrature components of the two drives. 

By analytically calculating the gradient of the fidelity with respect to all the control fields, the optimization problem can be efficiently solved by directly using the quasi-Newton optimization algorithms. In practice, we also add some additional penalty terms to the optimization cost function, in order to make the resulting solution robust to experimental imperfections.

\section{Frequency comb control method}\label{theory}
\subsection{Theory}
In the QEC experiment with a binomially encoded logical qubit in the storage cavity, the error syndrome is measured by mapping the photon number parity of the cavity state to the auxiliary qubit state. This is achieved by applying a classical microwave pulse containing $2M$ frequency components on the auxiliary qubit. In the interaction picture, the system dynamics are governed by the Hamiltonian:
\begin{eqnarray}
\label{H3}
\hat{H}_{I} = -\chi_\mathrm{qc}\hat{a}_\mathrm{c}^\dag\hat{a}_\mathrm{c}\ket{e}\bra{e}+ \sum_{n=1}^{2M}{\Omega_n e^{-i\delta_n t}\ket{e}\bra{g}+h.c.},
\end{eqnarray}
where $\Omega_n$ and $\delta_n$ are the drive amplitude and frequency detuning of the $n$-th driving component. For simplicity, $\chi_\mathrm{qc}$ will be denoted as $\chi$ in the following description. With the choice of $\delta_n = (2M-2n-1)\chi$ and $\Omega_n = \Omega$, each of the transitions of $|g, 2m+1\rangle \leftrightarrow |e, 2m+1\rangle$ with $m=0,1,2...M$ is resonantly driven by a microwave frequency component, and off-resonantly driven by $2M-1$ components with detunings of $2k\chi$ with $k=-(M-m)$ to $M+m-1$ and $k\neq 0$. On the other hand, each of the transitions $|g, 2m\rangle \leftrightarrow |e, 2m\rangle$ is also off-resonantly driven by $2M$ frequency components with detunings of $(2k-1)\chi$ with $k=-(M-m)$ to $M+m-1$.

For the cavity in the code space with a 2-photon state, the transition $|g, 2\rangle \leftrightarrow |e, 2\rangle$ is driven by $M$ pairs of frequency components with detunings $\pm\chi, \pm3\chi,...\pm (2M-1)\chi$. After a pulse duration of $T$, the initial state $|g, 2\rangle$ evolves to 
\begin{eqnarray}
\label{H5}
\cos \xi \ket{g,2}- ie^{2i\chi_{qc}T}\sin \xi \ket{e,2},
\end{eqnarray}
where 
\begin{eqnarray}
\xi = \sum_{n=0}^{M-1}2\int_{0}^{T}{dt \Omega \cos\left[ (2n+1)\chi t \right]}.
\end{eqnarray}

For $\Omega$ being a constant, we have $\xi = 2\sum_{n=0}^{M-1}\frac{\Omega}{(2n+1)\chi}\sin\left[(2n+1)\chi T\right]$. With the choice of $\chi T = m \pi$ ($m = 1, 2, 3, ...$ is an integer), the qubit finally returns to the ground state and nothing changes. 

For the cavity in the code space with 0- and 4-photon states, the auxiliary qubit's $|g\rangle \leftrightarrow |e\rangle$ transition is driven by $M-1$ pairs of frequency components with detunings of $\pm \chi, \pm 3\chi, ..., \pm (2M-3)\chi$ and two unpaired frequency components with detunings of $(2M\pm 1)\chi$. Supposing that $(2M-1)\chi \gg \Omega$, the effect of the two unpaired frequency components can be neglected. After the pulse duration $T$, the initial states $|g,0\rangle$ and $|g,4\rangle$ make a cyclic evolution and return to the original states. 

For the cavity in the error space with 1- and 3-photon states, the auxiliary qubit's transition $|g\rangle \leftrightarrow |e\rangle$ is driven by $M-1$ pairs of frequency components with detunings of $\pm 2\chi, \pm 4\chi,...\pm 2(M-1)\chi$, one resonant frequency component, and one unpaired frequency component with detuning of $2M\chi$. When the effect of the unpaired drive is neglected under the condition of $2M\chi \gg \Omega$, the system's evolution is
\begin{eqnarray}
\label{H7}
\cos \mu \ket{g,k}- ie^{ik\chi T}\sin \mu \ket{e,k}, k=1,3,
\end{eqnarray}
where 
\begin{eqnarray}
\label{H8}
\mu = \Omega T + 2\sum_{n=1}^{M-1} { \frac{\Omega}{2n\chi} \sin (2n\chi T)}.
\end{eqnarray}

With the choice of $\Omega T = \pi/2$ and $\chi T = m \pi$ with $m=1,2,3,...$ being an integer, both the $\ket{g,1}$ and $\ket{g,3}$ states are transformed to $\ket{e,1}$ and $\ket{e,3}$, respectively. Therefore, the detection of the auxiliary qubit in state $\ket{g}$ indicates that no photon loss has occurred, and the logical qubit remains in the code space with even parity. The detection of the auxiliary qubit in $\ket{e}$ indicates that a single-photon-loss error has occurred and the cavity is in the error space with odd parity.

\begin{figure*}[tb]
  \centering
  \includegraphics{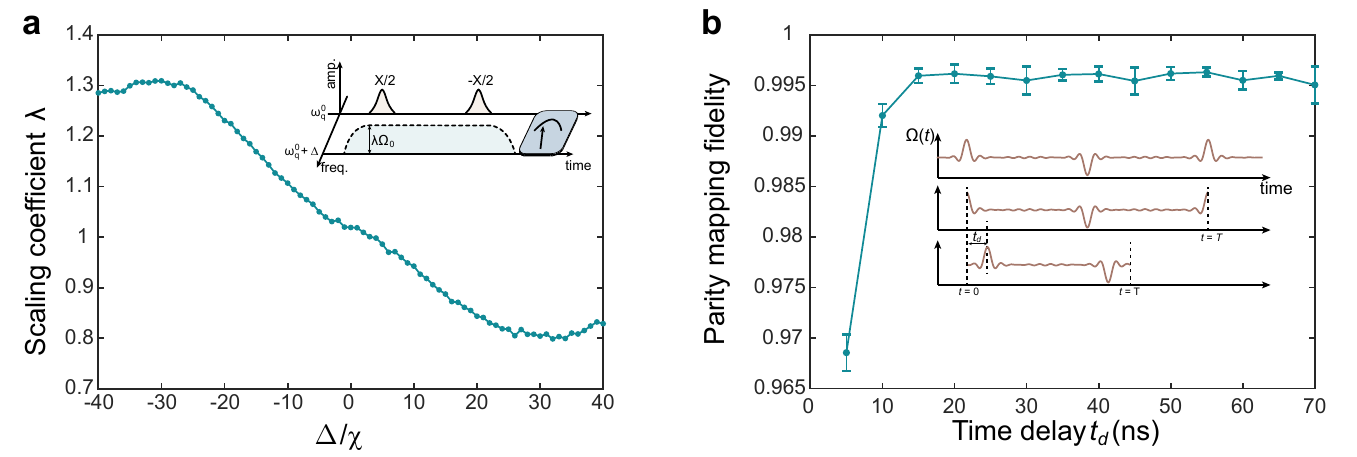}
  \caption{Calibration and optimization of the frequency comb pulse. (a) Calibration of the scaling coefficient of the pulse amplitude for each frequency component. (b) The measured parity detection fidelity as a function of the delay time of the frequency comb pulse envelope in order to suppress the pulse distortion. }
  \label{fig_comb_cal}
\end{figure*}

\subsection{Frequency comb pulse optimization}
According to the theory described above, all frequency components are assumed to have identical driving amplitudes, which are chosen to be $\Omega=\chi/4$ in our experiment for the parity mapping. However, the auxiliary qubit may have different responses for each frequency component drive with different frequency detunings due to the frequency-dependent transmission characteristic of the microwave control circuit. Thus, to achieve high-fidelity parity measurement, the pulse amplitude of each frequency component should be carefully calibrated.  

The drive amplitude of each frequency component is calibrated by performing a Ramsey experiment and measuring the a.c. Stark frequency shift. The magnitude of the a.c. Stark shift of the qubit frequency can be expressed as
\begin{eqnarray}
\label{H9}
| \delta | = \frac{1}{2} \left( \sqrt{\Delta^2+(\lambda\Omega_0)^2}-|\Delta| \right),
\end{eqnarray}
where $\Delta$ is the frequency detuning of the microwave tone, $\Omega_0$ is the uncalibrated drive strength of the detuned pulse, and $\lambda$ is the amplitude scaling coefficient. In our experiment, we calibrate the amplitude scaling by applying an off-resonant drive to the auxiliary qubit with frequency detunings of $\Delta = n\chi$ ($n=\pm 1, \pm 2,...$) and measuring the corresponding frequency shift from the Ramsey experiment. The calibration results are shown in Fig.~\ref{fig_comb_cal}a. 

In our experiment, the frequency comb pulse is generated by mixing the multi-frequency pulses with a single local oscillator, whose frequency is aligned to the qubit frequency with two photons in the cavity. Thus, the envelope of the frequency comb waveforms can be expressed as 
\begin{eqnarray}
\Omega_\mathrm{comb}(t) = \Omega\sum_{n=1}^{M}{\left\{\cos{\left[(2n-1)\chi (t - t_d) \right]}+\cos{\left[-(2n-1)\chi (t - t_d) \right]}\right\}},\quad t\in [0,T],
\end{eqnarray}
where $t_d$ is the delay time to introduce a phase shift for each frequency component. With $t_d=0$, the comb pulse envelope has a large amplitude at the initial and final time for the parity mapping, which will inevitably result in pulse distortions and reduce the control fidelity in the actual experiment. 

In our experiment, we introduce a phase shift for each frequency component by adding a delay time for the comb pulse envelope, thus making the pulse amplitudes approach zero at the initial and final time due to the destructive interference of all components. In the meantime, the total length of the parity mapping procedure can also be reduced by a factor of 2, approaching a time of about $\pi/\chi$ when choosing the drive strength $\Omega=\chi/4$. Therefore, the pulse duration for comb parity mapping is similar to that in the Ramsey interferometer but with a smaller pulse amplitude. In our experiment, we measure the parity fidelity as a function of the delay time, with the experimental result shown in Fig.~\ref{fig_comb_cal}b. An optimal value of the delay time is chosen as $47$~ns in our experiment to make the pulse amplitudes sufficiently small at the initial and final time. In addition, 5-ns rising and falling edges are also added to further smooth the waveforms of the frequency comb pulses. As a result, the optimal comb driving pulse for the parity measurement has a total length of $255$~ns with 22 frequency components in the QEC experiment.

\subsection{Parity measurement fidelity}
We first characterize the parity measurement fidelity by directly measuring the photon number parity of the vacuum state $\ket{0}$, giving a fidelity of 0.994. In addition, we also measure the photon number parities of even and odd cat states with different average photon numbers, which are generated by performing a parity measurement on an initial coherent state, post-selecting the parity measurement result, and performing another two consecutive parity measurements. The first two consecutive identical parity results would give a photon state parity with good confidence, and are post-selected to estimate the parity fidelity from the third parity measurement. The experiment gives an average parity fidelity of 0.987 for $\bar{n}=1$, 0.985 for $\bar{n}=2$, and 0.976 for $\bar{n}=3$.

For the logical qubit encoded with the lowest-order binomial code, the parity measurement fidelity for the cavity states in the code and error spaces are also measured in a similar manner. In our experiment, we first encode the cavity into each cardinal-point state in the code and error spaces, and then perform three consecutive parity measurements. Post-selection of the first two identical parity results would give better confidence for estimating the parity measurement fidelity from the third parity measurement. The experimental results are shown in Fig.~2 in the main text, indicating an average parity detection fidelity of 0.989 and 0.975 for the cavity states in the code and error spaces, respectively.

\begin{figure*}[tb]
  \centering
  \includegraphics{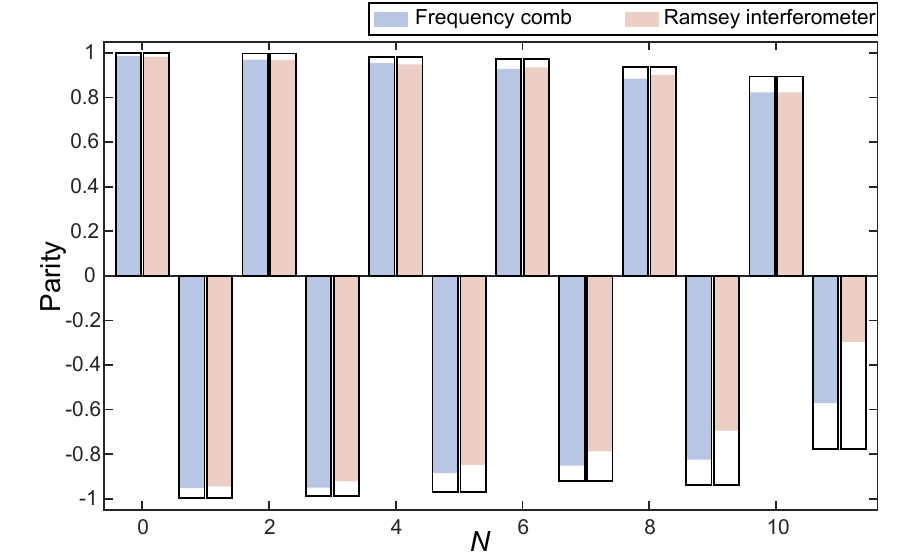}
  \caption{Bar chats of the measured parities for different photon number states $\ket{N}$ with both the frequency comb method and the Ramsey interferometer method. Solid black frames represent the ideal photon number parities from numerical simulations.}
  \label{fig_fockN}
\end{figure*}

\subsection{Comparison with the Ramsey method}
The conventional photon number parity detection of the cavity states is implemented by a Ramsey interferometer, where a qubit-state-dependent $\pi$-phase shift of the cavity is sandwiched between two unconditional $\pi/2$ pulses applied on the auxiliary qubit. During the parity mapping, the auxiliary qubit evolves in the equatorial plane of the qubit Bloch sphere most of the time and has an average excited state population of 0.5, no matter whether the cavity is in the code or error spaces. A qubit relaxation error during the parity mapping will give a wrong indication of the following correction operation, resulting in a depolarization error of the logical qubit. Meanwhile, the auxiliary qubit also suffers largely from dephasing noise during parity mapping with this type of parity measurement. In addition, the Ramsey interferometer necessitates unconditional $\pi/2$ pulses, which cannot be perfectly achieved for the multiphoton encoded logical states due to the photon-number-dependent dispersive shift of the auxiliary qubit frequency. Furthermore, it will generally deteriorate for large photon number encodings in the cavity, due to the inevitable introduction of off-resonant driving errors.

\begin{figure*}[tb]
  \centering
  \includegraphics{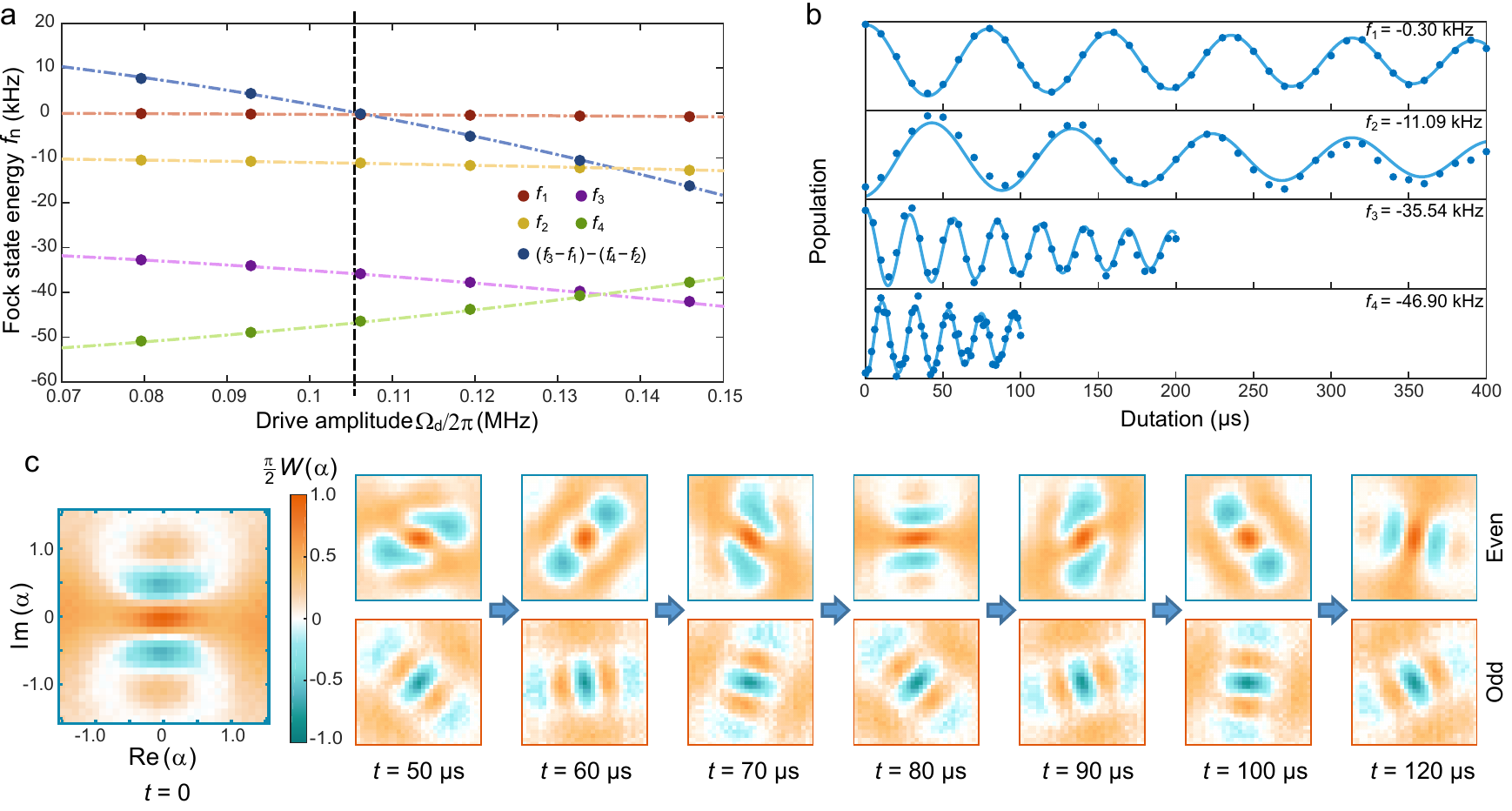}
  \caption{Eliminating the dephasing effect with PASS drive. (a) Measured phase accumulation rates $f_n$ for Fock states $\ket{n}$ (with $n=1,2,3,4$) relative to $\ket{0}$, as well as the difference between them $\Delta f = (f_3-f_1) - (f_4-f_2)$. (b) Measured Ramsey oscillations between Fock state $\ket{n}$ (with $n=1,2,3,4$) and $\ket{0}$ at the optimal amplitude of the PASS drive. (c) Measured Wigner functions of the cavity states after different evolution times conditional on the parity measurement results. }
  \label{fig_Kerrcancel}
\end{figure*}

As a distinct contrast, the ingenious designed frequency comb method for the parity mapping procedure in QEC can mitigate these adverse effects to some extent. Compared to the Ramsey method, the auxiliary qubit would have a small excited state population (less than 0.5) during the frequency comb parity mapping procedure for the cavity in code space with 0-, 2- and 4-photon states with carefully designed frequency comb parameters. In a QEC experiment, there is only a small probability of the single-photon-loss error occurring and the cavity still remains in the code space with a large proportion. Therefore, reducing the auxiliary qubit excited state population when the cavity is in the even parity subspace will suppress the auxiliary qubit relaxation errors during the parity mapping and thus finally benefit the whole QEC process. Besides, the continuous driving comb effectively decouples the auxiliary qubit from dephasing noises during the error syndrome mapping~\cite{guo2018}.

In addition, this frequency comb method also allows for measuring the parity of large photon number states because of no need for unconditional qubit rotations. In our experiment, we directly compare the performance of the frequency comb method with that of the conventional Ramsey method to measure the parities of large photon number states. In Fig.~\ref{fig_fockN}, we present the measured photon number parities with these two methods for various initial Fock states $\ket{N}$, which are generated by using the numerical optimization method. The experimental results indicate that the comb method indeed has an obvious benefit for large photon number states.

As another distinct advantage, the frequency comb control method can also be generalized to perform error syndrome detection of higher-order binomial codes, which can be used to detect and correct the errors for more than one photon losses~\cite{michael2016}. The basis states of these codewords have different photon number parities, with the generalized photon number parity serving as the error syndrome, which cannot be directly measured with the conventional Ramsey method. In contrast, the high tailorability of the frequency comb method does not require that the two basis states of the codeword have the same parity and can be easily adapted to the higher-order binomial codes.

\section{Details of the QEC Procedure}

The QEC experiment starts by initializing both the auxiliary qubit and the storage cavity in the ground state $\ket{g,0}$, which is achieved by post-selections of the auxiliary qubit's ground state and a subsequent cavity parity measurement since both the auxiliary qubit and the storage cavity have small thermal excitations.  

After the initialization, the auxiliary qubit is first prepared to the six cardinal-point states in the qubit Bloch sphere, and transferred to the binomial logical space \{$\ket{0_\mathrm{L}}$, $\ket{1_\mathrm{L}}$\} of the storage cavity with an encoding process, which is realized by applying a numerically optimized pulse with a duration of about 770~ns. 

The quantum information stored in the logical states is protected by repetitive QECs implemented after the encoding operation, and is extracted by decoding back to the auxiliary qubit and performing the tomography experiment on the auxiliary qubit. A single QEC cycle has a total duration of about 92.46~$\mu$s containing: a waiting time of $t_w \approx 90~\mu$s, a frequency comb pulse (with a total width of about 255~ns) to measure the photon number parity, a qubit readout pulse with a duration of about $600$~ns, a waiting time of about $511$~ns after the measurement to release the readout cavity photons and demodulate and digitize readout signals for feedback control, a $20$~ns cosine-shaped unconditional $\pi$ pulse to reset the auxiliary qubit, and correction operations (GRAPE pulse with a width of about 770~ns) conditional on the previous measurement result.

During the waiting time of $t_w$ in each QEC cycle, an off-resonant drive pulse with smooth rising and falling edges (100~ns for each) is applied on the auxiliary qubit to mitigate the self-Kerr induced dephasing effect of the logical codewords in the storage cavity by using the photon-number-resolved a.c. Stark shift (PASS) method~\cite{ma2020}. In our experiment, we measure the phase accumulation rates $f_n$ for Fock states $\ket{n}$ with $n=1,2,3,4$ relative to the vacuum state $\ket{0}$ as a function of the drive amplitude but with a fixed frequency detuning of $-3.5\chi$, and present the results in Fig~\ref{fig_Kerrcancel}a. By adopting an optimal drive amplitude of $\Omega_d/2\pi=0.106$~MHz, the accumulation rates measured from Ramsey experiments and shown in Fig.~\ref{fig_Kerrcancel}b, meet the error-transparent condition of $(f_4-f_2) - (f_3-f_1) = 0$ to eliminate the dephasing effect of the logical qubit. In order to further check the quantum evolution in the code and error spaces, we measure the Wigner functions of the cavity states by post-selecting the parity measurement result after various evolution time, with the experimental results shown in Fig.~\ref{fig_Kerrcancel}c. The results indicate that the phase coherence in the error space is significantly preserved, manifesting the tolerance of the stochastic single-photon-jump error during the waiting time.

\begin{figure*}[tb]
  \centering
  \includegraphics{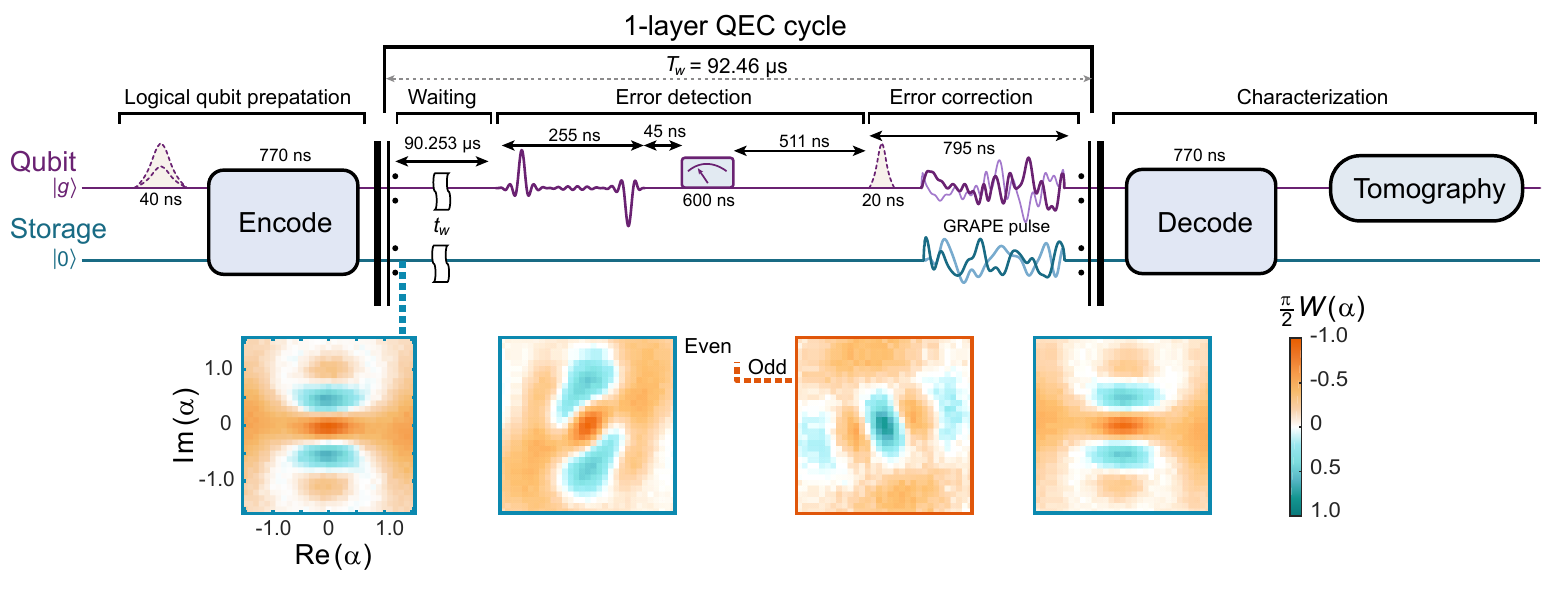}
  \caption{Experimental sequence for the one-layer QEC experiment, as well as the Wigner snapshots of the cavity state at each time step in the QEC sequence.}
  \label{fig_1layer_qec}
\end{figure*}

Figure~\ref{fig_1layer_qec} shows the experimental sequence of the one-layer QEC process, as well as the Wigner snapshots at each time step in the sequence. The feedback latency of the adaptive control is defined as the time interval between sending out the last point of the readout signal and sending out the first point of the qubit control signal, which also includes the travelling time through the whole experimental circuitry, and is about 511~ns in our experiment.  

\begin{figure*}[b]
  \centering
  \includegraphics{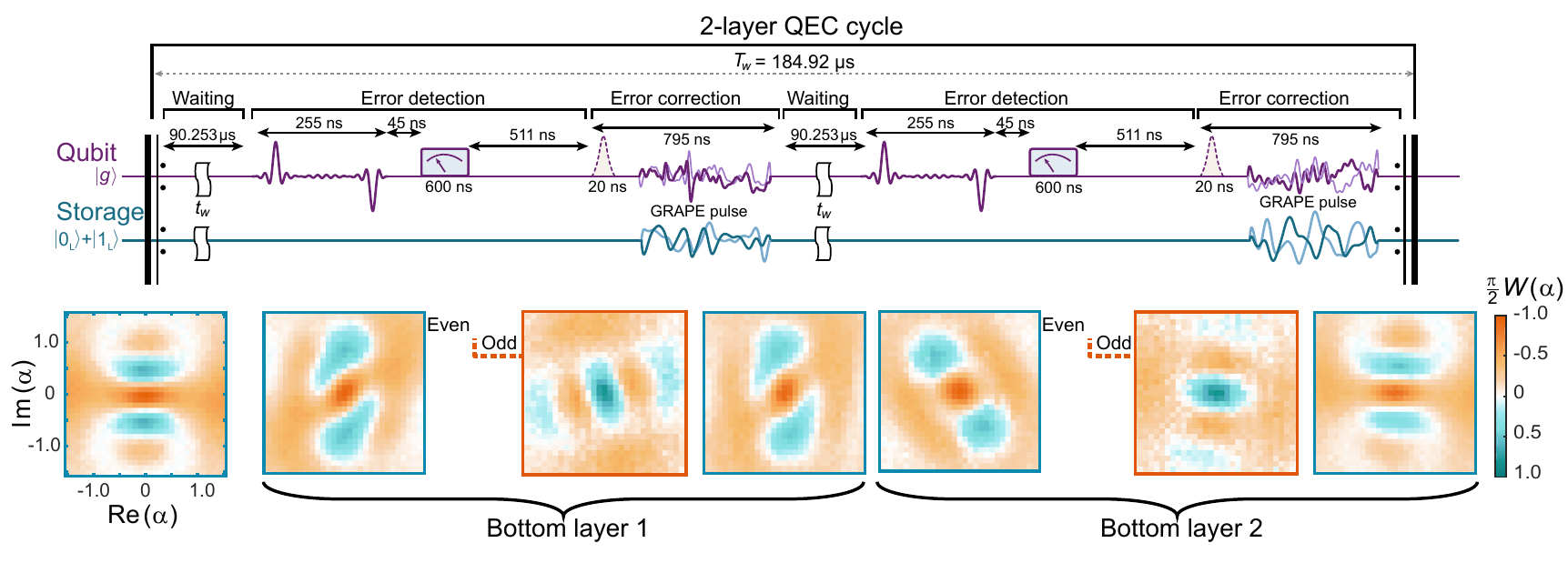}
  \caption{Experimental sequence for the two-layer QEC experiment, as well as the Wigner snapshots of the cavity state at each time step in the QEC sequence.}
  \label{fig_2layer_qec}
\end{figure*}

In order to balance the operation errors, no-parity-jump backaction errors, and photon-loss errors, we adapt a two-layer QEC procedure~\cite{hu2019} to improve the error correction performance, with the protocol details shown in Fig.~\ref{fig_2layer_qec}. 

\section{Error analysis}

\begin{figure*}[tb]
  \centering
  \includegraphics{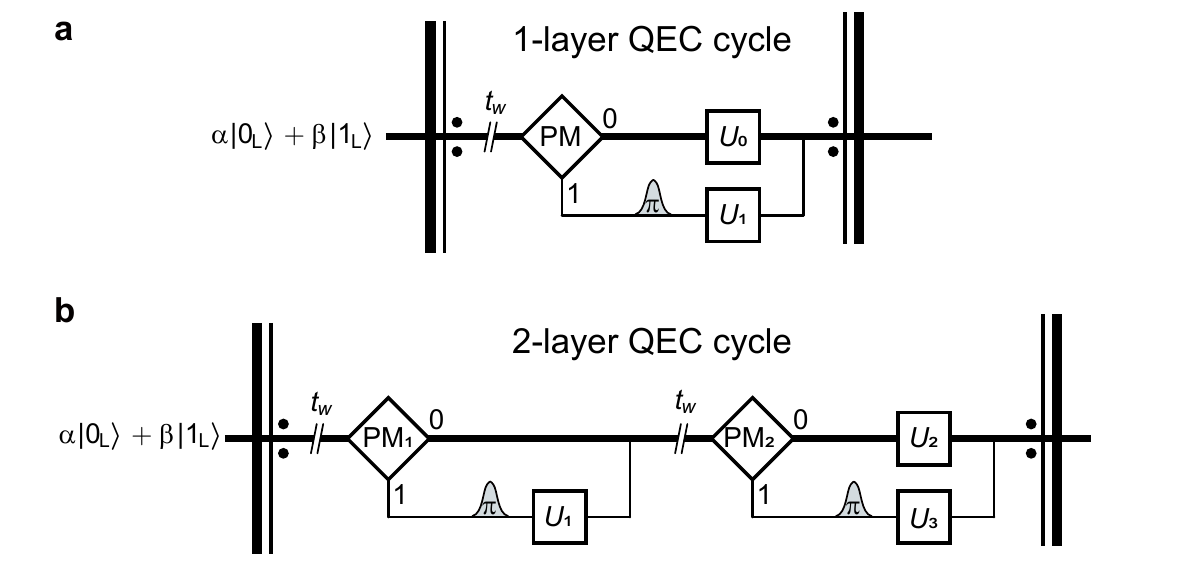}
  \caption{Schematic of the one-layer (a) and two-layer (b) QEC cycles in the experiment. }
  \label{fig_qec_schematic}
\end{figure*}

In order to understand the experimental results of the QEC performance, we investigate the error sources and their contributions to the loss of fidelity for each QEC cycle in this section. According to an analytical model in Ref.~\cite{hu2019}, we divide the error sources in the QEC procedure into four parts: the intrinsic error, the parity measurement infidelity, the recovery operation infidelity, and the auxiliary qubit thermal excitation error for both the one- and two-layer QEC experiments, with the schematics shown in Fig.~\ref{fig_qec_schematic}. These errors are summarized and listed in Table~\ref{Table_error}. Detailed descriptions for estimating and calculating these errors are presented below. Note that the process fidelity has a minimum of 0.25, and thus in the following we define the fidelity as the normalized process fidelity $F = (F_\chi - 0.25)/0.75$ with a full scale between 0 and 1.

\begin{table}[b]
\caption{Error budget for the one- and two-layer QEC processes. *These errors are estimated from numerical simulations.}
\centering
\begin{tabular}{p{2cm}<{\centering}  p{1.5cm}<{\centering} p{1.5cm}<{\centering} p{1.5cm}<{\centering} p{1.5cm}<{\centering} p{1.5cm}<{\centering} p{1.5cm}<{\centering} p{1.5cm}<{\centering} p{1.5cm}<{\centering}}
  \hline
  \hline
  \multicolumn{2}{c}{Parameters} &  Intrinsic error* & Detection error & Recovery error & Thermal error & Average error & Predicted lifetime & Measured lifetime \\\hline
  &&&\\[-0.9em]
  \multirow{2}{*}{1-layer QEC} & case 0 &  6.7\% & 1.1\% & 2.7\%  & 0.8\% & \multirow{2}{*}{ 11.5\% } & \multirow{2}{*}{ 757~$\mu$s } & \multirow{2}{*}{ 755$\pm 9~\mu$s } \\
                               & case 1 &  5.3\% & 2.5\% &  3.9\% & 0.8\% & \\
  &&&\\[-0.9em]
  \hline
  &&&\\[-0.9em]
  \multirow{4}{*}{2-layer QEC} & case 00 &  12.1\% & 2.2\% & 2.7\%  & 1.1\% & \multirow{4}{*}{ 20.1\% } & \multirow{4}{*}{ 824~$\mu$s } & \multirow{4}{*}{ 805$\pm 18~\mu$s } \\
                               & case 01 &  14.8\% & 3.6\% &  3.9\% & 1.1\% & \\
                               & case 10 &  9.2\% & 3.6\% &  6.6\% & 1.1\% & \\
                               & case 11 &  20.4\% & 5.0\% &  7.8\% & 1.1\% & \\
  &&&\\[-0.9em]
  \hline 
  \hline
\end{tabular}
\label{Table_error}
\end{table}

1. The intrinsic error comes from the fact that the lowest-order binomial code implemented in this experiment can only protect quantum information against a single-photon-loss error, therefore the errors that are not in the error set \{$\hat{I}$, $\hat{a_c}$\} would cause a failure of the QEC operation. These errors include multiple-photon-loss errors, the photon gain error due to thermal excitations, the dephasing errors due to the combination of photon jumps and the self-Kerr effect, and the non-unitary no-jump evolution. Note that the self-Kerr effect can be eliminated by the PASS method as previously discussed. In order to quantify the contribution of these errors, we perform numerical simulations of the QEC process with including only the cavity decay and Kerr terms but without the auxiliary qubit decoherence. In the simulation, we also add PASS drives to include Kerr cancellation and consider the cavity relaxation errors during the error detection in each QEC cycle. Besides, we also add an auxiliary qubit excitation error of about 1\% for each PASS drive to include the PASS-induced excitation errors~\cite{ma2020}. For the one-layer QEC experiment, the simulation gives an intrinsic error of about $\epsilon_{i0}=6.7\%$ and $\epsilon_{i1}=5.3\%$ for case 0 (detecting no error with a probability $p_0=0.781$) and case 1 (detecting one error with a probability $p_1=0.219$), respectively. The simulation of the two-layer QEC experiment gives an intrinsic error of about $\epsilon_{i00}=12.1\%$, $\epsilon_{i01}=14.8\%$, $\epsilon_{i10}=9.2\%$, and $\epsilon_{i11}=20.4\%$ for case 00 (both detecting no error with a probability $p_{00}=0.630$), case 01 (first detecting no error and second detecting one error with a probability $p_{01}=0.152$), case 10 (first detecting one error and second detecting no error with a probability $p_{10}=0.174$), and case 11 (both detecting one error with a probability $p_{11}=0.044$), respectively.

2. The parity measurement infidelity is quantified by directly measuring the photon number parities of the cavity states encoded in the six cardinal-point states in the code and error spaces with three consecutive frequency comb parity measurements. The experimental results are shown in Fig.~2(b) in the main text, indicating an average infidelity of $\epsilon_{D_0} = 1.1\%$ and $\epsilon_{D_1} = 2.5\%$ for the cavity states in these two spaces, respectively.

3. The recovery gates for correcting both the single-photon-loss error and no-photon-jump error are implemented by using numerically optimized pulses with the GRAPE method. It is difficult to experimentally calibrate these GRAPE pulses directly because both the initial state preparation and final state measurement also require GRAPE pulses for the encoding and decoding operations. Since the optimization procedure, pulse duration, and hardware are all the same, we assume that the fidelities of all the GRAPE pulses are the same. The gate infidelity of each GRAPE pulse can be roughly estimated from the encoding-decoding process, with a value of $\epsilon_{U_0} = \epsilon_{U_1} = \epsilon_{U_2} = \epsilon_{U_3} = 2.7\%$. After detecting the auxiliary qubit in the excited state, an unconditional $\pi$ pulse is applied to reset the qubit before the GRAPE operations and results in an error of about $0.7\%$ from the qubit process fidelity with zero idling time. Note that the auxiliary qubit also has a relaxation error of about $0.5\%$ during the feedback latency when detected in the excited state. Thus these two contributions give a total error of about $\epsilon_\pi = 1.2\%$ for the reset $\pi$ pulse.

4. The auxiliary qubit thermal excitation errors come from the small probability of the excitation of the auxiliary qubit to the $\ket{e}$ state during the total cycle time of about $T_w \approx 92~\mu$s and $T_w \approx 184~\mu$s for one- and two-layer QEC, respectively. The estimation from $\epsilon_\mathrm{th} = n_\mathrm{th}^\mathrm{q} (1 - e^{-T_w/T_1^\mathrm{q}})$ gives average errors of 0.8\% and 1.1\% for one- and two-layer QEC, respectively. 

\begin{figure*}[tb]
  \centering
  \includegraphics{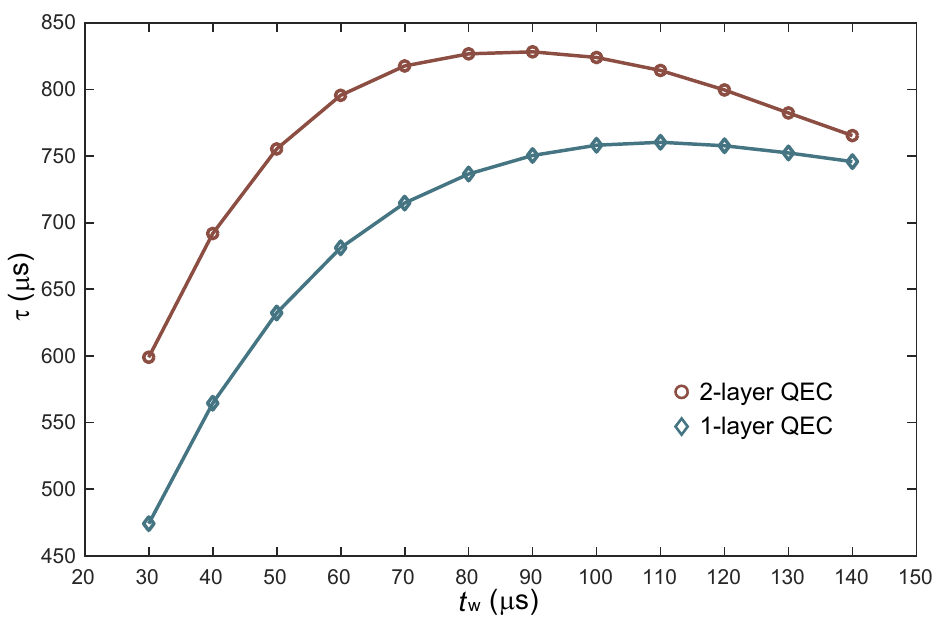}
  \caption{The numerically estimated lifetimes of both one- and two-layer QECs as a function of the waiting time in each QEC cycle. }
  \label{fig_lifetime}
\end{figure*}

With all these errors for individual operations in hand, we can calculate the weighted average total errors from the following equation:
\begin{eqnarray}
\epsilon_1 &=& p_0\left(\epsilon_{i0} + \epsilon_{D_0} + \epsilon_{U_0} + \epsilon_\mathrm{th} \right)  \notag\\
           &+& p_1\left(\epsilon_{i1} + \epsilon_{D_1} + \epsilon_{U_1} + \epsilon_\pi + \epsilon_\mathrm{th}\right),
\end{eqnarray}
for the one-layer QEC experiment, and the following equation:
\begin{eqnarray}
\epsilon_2 &=& p_{00}\left(\epsilon_{i00} + \epsilon_{D_0} + \epsilon_{D_0} + \epsilon_{U_2} + \epsilon_\mathrm{th} \right)  \notag\\
           &+& p_{01}\left(\epsilon_{i01} + \epsilon_{D_0} + \epsilon_{D_1} + \epsilon_\pi + \epsilon_{U_3} + \epsilon_\mathrm{th} \right)  \notag\\
           &+& p_{10}\left(\epsilon_{i10} + \epsilon_{D_1} + \epsilon_\pi + \epsilon_{U_1} + \epsilon_{D_0} + \epsilon_{U_2} + \epsilon_\mathrm{th} \right)  \notag\\
           &+& p_{11}\left(\epsilon_{i11} + \epsilon_{D_1} + \epsilon_\pi + \epsilon_{U_1} + \epsilon_{D_1} + \epsilon_\pi + \epsilon_{U_3} + \epsilon_\mathrm{th} \right), 
\end{eqnarray}
for the two-layer QEC experiment.

According to a single exponential decay of the QEC process fidelity, we can estimate the decay time $\tau$ of the QEC process by
\begin{eqnarray}
\tau = -\frac{T_w}{\ln{\left(1-\epsilon \right)}}.
\end{eqnarray}
The predicted lifetimes for both the one- and two-layer QEC experiments are listed in Table~\ref{Table_error}, as well as the measured lifetimes, which are consistent with each other.

In addition, we also calculate the expected lifetimes of the one- and two-layer QECs as a function of the waiting time of the idle operation, and the results are shown in Fig.~\ref{fig_lifetime}. In our QEC experiments, we choose an optimal waiting time of about 90~$\mu$s to achieve the optimal QEC performance.

%